\pdfoutput=1
\documentclass[pra,twocolumn,showpacs]{revtex4}

\usepackage{graphicx}
\usepackage{epstopdf}
\usepackage{xcolor}

\usepackage{amsmath}
\usepackage{amssymb}
\usepackage{graphicx}
\usepackage{dcolumn}
\usepackage{bm}

\begin{document}

\preprint{APS/123-QED}

\title{Ballistic atom pumps}

\author{Megan K. Ivory$^{1}$, Tommy A. Byrd$^{1}$, Andrew J. Pyle$^{1}$, \\Kunal K. Das$^{2}$, Kevin A. Mitchell$^{3}$, Seth Aubin$^{1}$, and John B. Delos$^{1}$}

\affiliation{$^{1}$ Department of Physics, College of William and Mary, Williamsburg, VA 23187, USA }
\affiliation{$^{2}$ Department of Physical Sciences, Kutztown University of Pennsylvania, Kutztown, PA 19530, USA }
\affiliation{$^{3}$ School of Natural Sciences, University of California, Merced, CA 95344, USA}

\date{\today}

\begin{abstract}
We examine a classically-chaotic system consisting of two reservoirs of particles connected by a channel containing oscillating potential-energy barriers. We investigate whether such a system can preferentially pump particles from one reservoir to the other, a process often called ``quantum pumping.'' We show how to make a ``particle diode'' which under specified conditions permits net particle pumping in only one direction. Then we examine systems having symmetric barriers. We find that if all initial particle energies are considered, a system with symmetric barriers cannot preferentially pump particles. However, if only finite initial energy bands are considered, the system can create net particle transport in either direction. We study the system classically, semiclassically, and quantum mechanically, and find that the quantum description cannot be fully understood without the insight gained from classical and semiclassical analysis.
\end{abstract}

\maketitle

\section{\label{Introduction}Introduction}

A ballistic atom pump is a system containing two or more reservoirs of neutral
atoms or molecules and a junction connecting them containing a time-dependent
potential.  “Ballistic” means that atoms move through the pump as independent
particles. The theoretical description may be given by classical, semiclassical, or quantum
theories. Atom statistics may be Bose, Fermi, or Boltzmann.  We intend that the
definition of “ballistic atom pumps” be interpreted broadly (however, systems
having intrinsically many-body phenomena such as viscosity should not be called “ballistic;” if there are interactions among the particles, these interactions can be described by an average single-particle potential).

Particle transport is an ongoing topic of interest in a variety of systems from solid state circuitry to microfluidic devices to futuristic atomtronic components.  Since the advent of laser cooling, precise control and manipulation of neutral ultracold atoms has attracted attention to atomic systems that can mimic more challenging systems.  One such phenomenon in electronic solid-state systems describes electronic transport through mesojunctions \cite{Ferry-Goodnick} having time-dependent potential barriers, a phenomenon often called ``quantum pumping''\cite{GarttnerSchmelcher_PRE2010,brouwer-1,thouless}. The choice of potential likewise emulates the turnstile quantum pump usually studied in mesoscopic electronics \cite{turnstile,turn2}. Although quantum pumping has been theorized for decades \cite{thouless,brouwer-1,turnstile,turn2,mosk,kim,arrachea,avron2,blaa,samuelsson,RomeoCitro,Splettstoesser, RomeoCitro2}, there has only recently been an experimental realization of such a system due to the challenges of overcoming capacitive coupling and rectification effects in electronic systems \cite{Switkes_Science1999,Brouwer_PRB2001,josehpson-quantum-pump2}.  Recent proposals have suggested bypassing these difficulties by simulating a quantum pump in a system of neutral cold atoms \cite{Das2011,DasAubin_PRL2009}.

Neutral atom transport is also becoming increasingly important in its own right due to the ongoing development of atomtronics, which seeks to replicate critical tools of electronics in neutral atoms.  Analogues of batteries, diodes, transistors, and recently hysteresis \cite{zoz,pepinocooper,eckel} have been explored in ultracold neutral atom systems.  The motivation behind such devices is multifarious.  Unlike their electronic counterparts, these systems allow scientists to study analogous tools in well-controlled and idealized environments like optical lattices.  Additionally, long coherence times provide unique opportunities for quantum state preparation, storage, and readout, making atomtronic devices a serious competitor as a basis for quantum computers \cite{zhaochen}. Finally, neutral atoms present degrees of freedom not available in their electronic counterparts, such as bosons, fermions, and scalable interactions. In this paper, we present a detailed study of the classical and quantum features of a ballistic atom pump which has potential applications such as a battery, diode, or rectifier in atomtronic circuits.

The pumps we consider in this paper have two reservoirs and a pump which is
effectively one-dimensional, so the Hamiltonian is
\begin{eqnarray}
H(p,x) = p^2/2m + V(x,t).
\end{eqnarray}
We choose $V(x,t)$ to consist of two repulsive barriers oscillating with the
same frequency $\omega$, but not necessarily with the same amplitude or phase.
We study rectangular barriers (easiest theoretically) and Gaussian barriers
(easiest experimentally using optical forces).  The questions we address are:
Can such systems pump atoms preferentially from one side to the other without
an external bias, such as a difference in chemical potentials in the reservoirs?
In particular, can we make an atom “diode” that will allow atoms to pass
through the pump in only one direction? In order to understand the quantum features of
such a pump, it is necessary to develop a clear understanding of classical scattering by
a pair of oscillating potential barriers that function as a turnstile pump.

We begin with a precise specification of the models we study. Then we consider simple asymmetric pumps that rectify net particle transport, which we call "particle diodes" because they allow transport in only one direction for certain ranges of initial particle energy.
 These diodes have one barrier fixed and one oscillating barrier. Then we consider pumps that are symmetric in the sense that
the two barriers are
identical, but their oscillations are not in phase with each other.  We prove a
symmetry theorem which shows that such pumps can give no net particle pumping if the
behavior of the particles is classical and the initial phase-space distribution
is uniform in both reservoirs.  However, if the phase-space distribution is not
uniform, then such pumps can produce net particle transport in either direction. We also show that if
the two potential barriers are separated by a modest distance, atoms can get
stuck in a “complex” or ‘resonance zone” between them, and the system is a nice model of
chaotic transport \cite{ct1,ct2,ct3,ct4,ct5,ct6,ct7,ct8,ct9,ct10,ct11,ct12,ct13,ct14,ct15,ct16,ct17,ct18,ct19,ct20,ct21,ct22,ct23,ct24,ct25,ct26,ct27,ct28,ct29,ct30,ct31,sm1,km1,km2, km3,km4}.  (In a separate paper \cite{ByrdDelos} we have provided a topological description
of this chaotic transport).

The relationships among classical, semiclassical, and quantum descriptions for transport
past a single oscillating Gaussian barrier were discussed in detail in \cite{single}.
Consider the case that atoms enter the pump from one side with fixed momentum $p_i$ and
kinetic energy $E_i$.  In the quantum description, because the barriers are oscillating
with a fixed frequency, Floquet theory tells us that after passing through the pump, the
spectrum of transmitted energies is a set of narrow peaks at energies
$E_n=E_i +n \hbar \omega$, where $n$ is an integer.  The heights of these peaks can be computed numerically by
solving the Schroedinger equation, but in general no patterns are visible in those heights.

In the classical description (again assuming that particles enter with a fixed initial
momentum $p_i$ but a range of positions $x_i$), then the final momentum $p_f$ is a
bounded periodic function of the initial position  $x_i$,
$p_f = \mathcal{P}_f (x_i)$.  The upper and lower bounds of the range of this function define the
classically allowed region. Inside this classically allowed region, provided that $\mathcal{P}_f (x_i)$ is continuous, there must be
an even number of trajectories leading to each final momentum.  The distribution of final
momenta is a smooth function
except at extrema of $\mathcal{P}_f (x_i)$, where the distribution has an integrable
singularity.  One finds that the Floquet peaks obtained in the quantum description are large
primarily in the classically allowed region, with small spillover past the boundaries
(momentum-space tunneling).  Still the heights of peaks are incomprehensible.

Finally, in semiclassical theory, for each final momentum one sums over
the initial positions that give trajectories leading to that final momentum, and
incorporates phases for each such orbit (momentum-space action plus Maslov indices).
Summing over one cycle of $\mathcal{P}_f(x_i)$ produces a smooth function, and the
relative heights of the Floquet peaks are discrete values of it. Summing over many
cycles of $\mathcal{P}_f(x_i)$ causes the peaks seen in the quantum description to
emerge, with good agreement between the two methods (see Figs.~\ref{gaussrects} and~\ref{gauss2osc}). We show a few representative calculations of each type in this
paper, but we concentrate on the classical description, with the understanding
that semiclassical calculations can be carried out when desired, and that the semiclassical
description agrees well with the quantum description.

\section{\label{Model}Model}

Our atom pump consists of two repulsive potential barriers with amplitude oscillations
that have the same frequency, but are not in phase with one another:
\begin{eqnarray}
V(x,t) = U_L(x,t) + U_R(x,t)
\end{eqnarray}
In this paper we examine both rectangular
and Gaussian potentials.  The rectangular barrier potentials are given by:
\begin{subequations}
\begin{align}
U&_L(x,t)=
\left\{
     \begin{array}{cl}
          \hat{U}_L\left(1+\alpha_L \cos(\omega t)\right)&,\mbox{ }b_{L-}<x<b_{L+} \\
          0 &, \mbox{ elsewhere}
 \end{array}
   \right. \label{recleft}\\
U&_R(x,t)=\left\{
     \begin{array}{cl}
          \hat{U}_R\left(1+\alpha_R \cos(\omega t+\phi)\right)&, b_{R-}<x<b_{R+} \\
          0&, \mbox{elsewhere}
 \end{array}
   \right. ,\label{recright}
\end{align}
\end{subequations}
where $b_{L-}=-\hat{x}-\sigma_L$, $b_{L+}=-\hat{x} +\sigma_L$, $b_{R-}=\hat{x}-\sigma_R$, $b_{R+}=\hat{x} +\sigma_R$, $\hat{U}_{L,R}$  is the average height of each barrier, $\alpha_{L,R}$ is the amplitude of
oscillation of each, $\omega=2\pi/T$ is the common frequency and $T$ is the period, $\phi$
is an additional phase term, and $2\sigma_{L,R}$ is width of each barrier. The left and
right barriers are centered at $x=-\hat{x}$ and $x=\hat{x}$, respectively, and always have a center-to-center distance of $\Delta x = 2\hat{x}$. When the barriers touch, i.e., have no separation, $\sigma_{L,R} = \hat{x}$.
If only the left-hand barrier is oscillating then $\alpha_R = 0$.

\begin{figure}[t]
\includegraphics*[width=\columnwidth]{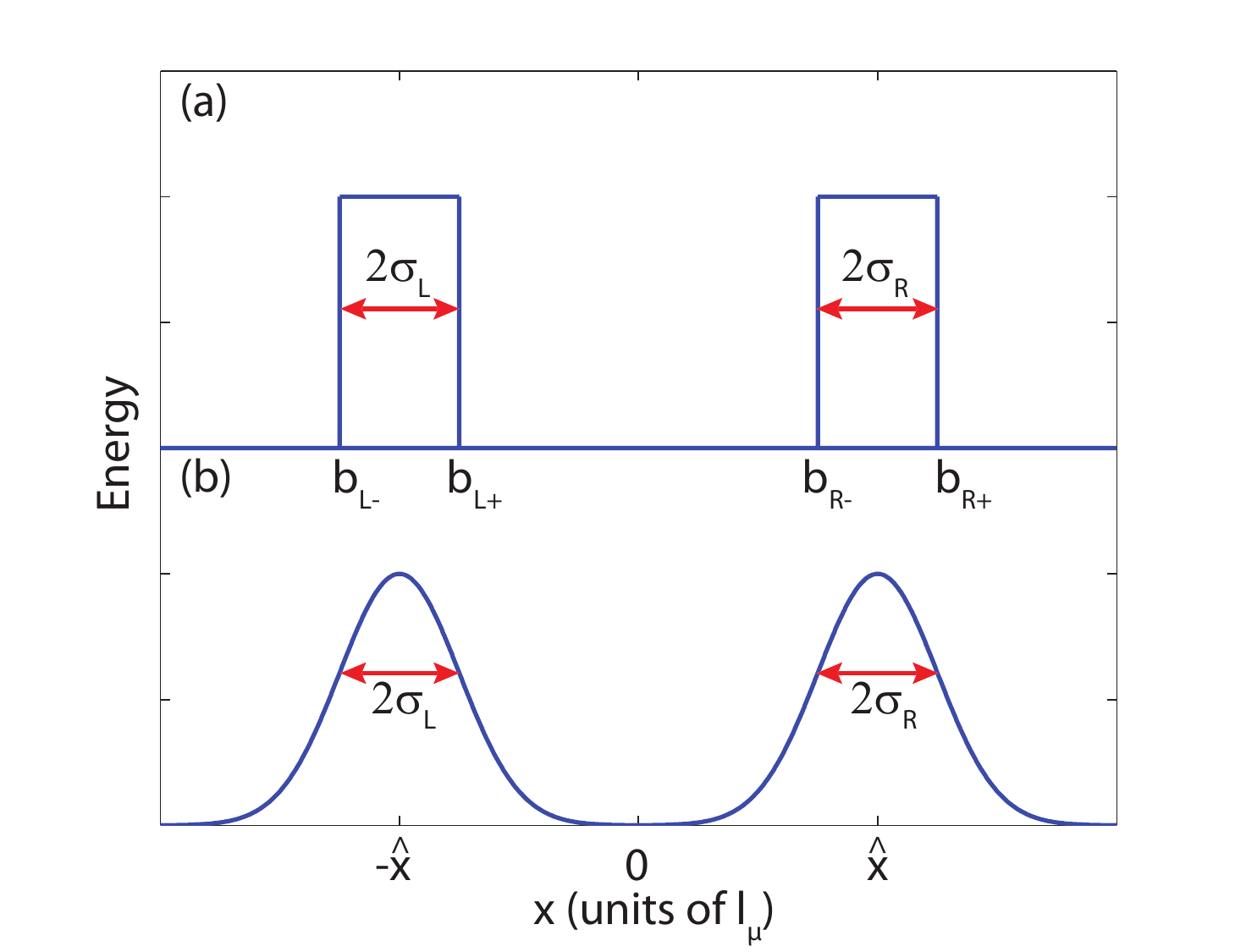}
\caption{(Color Online) Types of barriers considered in this paper. (a) Rectangular barriers, which are centered at $\pm\hat{x}$ and have width $2\sigma_{L,R}$. If $\sigma_{L,R}=\hat{x}$, the barriers touch. (b) Gaussian barriers centered at $\pm\hat{x}$ which have standard deviation $\sigma_{L,R}$. }\label{barrpar}
\end{figure}

The Gaussian potential barriers
are given by
\begin{subequations}
\begin{align}
U_L(x,t)&=\hat{U}_L \left(1+\alpha_L \cos(\omega t)\right)\exp\left(\frac{-(x+\hat{x})^2}{2\sigma_L^2}\right) \label{gaussleft} \\
\begin{split}
U_R(x,t)&=\hat{U}_R \left(1+\alpha_R \cos(\omega t+\phi)\right)\\
&\times\exp\left(\frac{-(x-\hat{x})^2}{2\sigma_R^2}\right) \label{gaussright},
\end{split}
\end{align}
\end{subequations}
where $\sigma_{L,R}$ is the standard deviation of the Gaussian. Fig.~\ref{barrpar} shows the parameters for the barriers.

Without loss of generality, we can choose units of mass, energy and time such that $m=1$, $\hat{U}_L = \hat{U}_R = 1$, and $\omega = 1$. The remaining parameters are the barrier widths $\sigma_L$ and $\sigma_R$, the barrier oscillation amplitudes $\alpha_L$ and $\alpha_R$, and the phase difference between the barriers, $\phi$.  In this paper, we typically choose $\sigma_L=\sigma_R$. In quantum and
semiclassical mechanics one additional parameter arises, the value of $\hbar$, which we set as $\hbar=1$. The general way to apply such scaling principles is given in \cite{scale}.

The units used in this paper are theoretical, and are the same as those used in \cite{single}. Namely, the choice of a theoretical unit convention based on $\hbar=1$ and $m=1$ is equivalent to selecting an arbitrary time unit $t_{u}$ and a related length unit $l_{u}=\sqrt{\hbar t_{u}/m}$,  with $\hbar=1.054 \times10^{-34}$ J$\cdot$s. The corresponding energy unit is $E_{u}=\hbar/t_{u}$, while the mass unit is that of the particle, $m_{u}=m$.

We start with a distribution of particles far to the left of the barriers, far to the right, or
both.  For our classical calculations, the distribution has a single momentum (i.e. it is a
delta-function in momentum space centered at $p_i$).  The distribution in position space is
uniform over a length $L=vT$  where $v$ is the initial velocity of the particles (i.e. uniform over
a length corresponding to the distance the incident particles travel in one cycle of the
barriers).  In semiclassical and quantum calculations, we begin with a wave packet that is
narrow in momentum space, centered at $p_i$, and correspondingly wide in position space,
$\Delta x_i >> L$.  Thus its magnitude is nearly uniform over the length $L$ corresponding to a cycle. The wave function in position space at the initial time is given by
\begin{equation}
\Psi(x_i,t_i=0)=F(x_i)e^{i p_ ix_i}
\end{equation} where $F(x_i)$ is
\begin{equation}
F(x_i)=\left(1/2\pi \right)^{1/4}e^{-(x_i + x_c)^2/4\beta^2}.\label{rectenv}
\end{equation}The initial probability density is thus $|\Psi(x_i,t_i=0)|^2=F^2(x_i)$, which is a Gaussian centered at $-x_c$ with
 standard deviation $\beta$. Our quantum calculations are performed in the same fashion as in \cite{single}, and are based on propagating the wave packet with the time-dependent Schr\"odinger equation via a split-step operator method \cite{G.P.Agarwal}.

We determine the net particle transport in these systems by the following process: 1) For each
initial momentum, launch particles  toward the barriers from the left, and compute and record
the fraction transmitted and reflected.  Also record the final momenta of transmitted and
reflected particles.  2) Launch particles with the same initial energy toward the barriers
from the right, and compute the fraction transmitted and reflected, and their final momenta.
3) Sum the results of each of these to obtain the net fraction of particles transmitted left
to right (which may be negative if more are transmitted from right to left). 4)  If appropriate, average over initial momenta.

We define the fractional transport of particles through the pump as
\begin{align}
C_P(|p_i|)=\frac{R(|p_i|)-L(|p_i|)}{R(|p_i|)+L(|p_i|)},\label{partcurr}
\end{align}
where $R(|p_i|)$ is the number of particles per cycle scattered to the right for each $|p_i|$, and $L(|p_i|)$ is the number of particles per cycle scattered to the left. The sum $R(|p_i|)+L(|p_i|)$ represents all particles for a given $|p_i|$. $C_P(|p_i|)$ is positive when more particles are scattered to the right for a given $|p_i|$, and negative when more particles are scattered to the left. When equal numbers of particles scatter to the right and left, e.g. when all particles are reflected or transmitted, $C_P(|p_i|)=0$.

\section{\label{Rectifiers}Particle Diodes}

\subsection{An Elevator Model}\label{RecElev}

A double-barrier particle pump can make a kind of diode, in which net particle pumping can only be in one direction if the initial particle energy is sufficiently small. This type of diode consists of one static barrier which is high enough to prevent transmission of particles incident from one direction, and one oscillating barrier which can lift particles approaching from the other direction over the static barrier. This is analogous to photon-assisted tunneling \cite{TienGordonPhysRev1963,Pimpale_JPhysA1991, FedirkoVyurkov_PhysicaStatusSolidiB2000, jauho}. When the incident energy of particles is greater than the height of the static barrier, net particle transport is only possible in the opposite direction. It is simplest if the two barriers are touching each other.  Let us simplify the description of the potentials to
\begin{align}
\begin{array}{cllrlll}
U_R(x,t)&= \hat{U}&,&   0&<x&<&b\\
U_L(x,t)&=Q(t)&,&  -b&<x&<&0,
\end{array}
\end{align}
where $Q(t)$ is a periodic function of $t$ with period $T=2 \pi$, and $\hat{U}$ is a constant.  Suppose
\begin{eqnarray}
Q(t)&= \left\{
\begin{array}{lllllll}
0&,&     0&<& t\pmod{2\pi}&<&\pi\\
\hat{U}&,&   \pi&\leq& t\pmod{2\pi}&<&2 \pi.
\end{array}
\right.
\end{eqnarray}

\begin{figure}[t]
\includegraphics*[width=\columnwidth]{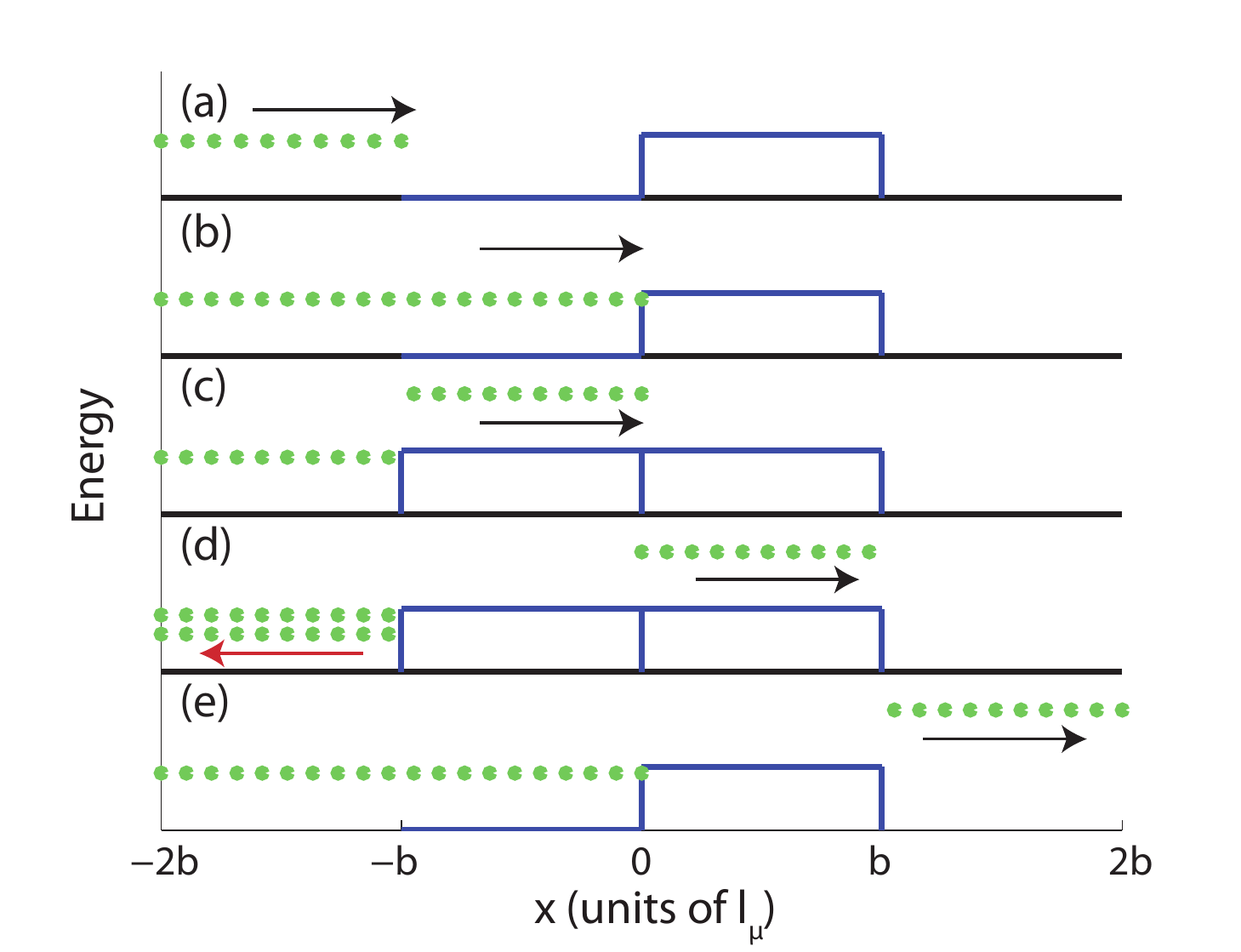}
\caption{(Color Online) Qualitative schematic of a diode with rectangular barriers. Particles approach from the left in (a) and (b). In (c), the left barrier abruptly rises to $E=\hat{U}$, and particles on top of it gain enough energy to transmit past the right barrier, as seen in (d) and (e).}\label{rectexample}
\end{figure}

Then particles incident from the right with kinetic energy $K_i<\hat{U}$ cannot pass over the right barrier.  From the left, (see Fig.~\ref{rectexample}(a)) a stream of particles having fixed kinetic energy $K_i<\hat{U}$ and density independent of position all file into the elevator when it is on the ground floor (Fig.~\ref{rectexample}(b)), and then at $t=\pi$ they are lifted abruptly to the penthouse level on the roof, where the back door of the elevator opens (Fig.~\ref{rectexample}(c)).  The particles keep their kinetic energy in this process, and politely file out in line onto the roof (Fig.~\ref{rectexample}(d)).  At $t=2\pi$ they are all lined up on top of the right-hand barrier, and one-by-one they slide down the edge of that barrier and escape to the right with kinetic energy $K_f = K_i + \hat{U}$ (Fig.~\ref{rectexample}(e)).  Meanwhile the door of the elevator has slammed again and it abruptly returns to ground level.

For this system, if the barrier width is $b=|p_i|\pi/\omega$ and if the first particle arrives at the left edge of the left barrier at $t=0$, half of the particles incident on the elevator from the left -- the ones that arrive for $0<t$(mod$2\pi)<\pi$ -- go over the barrier, and the other half -- arriving for $\pi \leq t$(mod$2\pi)<2\pi$ -- are reflected by the left-hand barrier, so the transmitted fraction is half of the incident fraction.

Is this the theoretical maximum for transport?  We cannot think of any other function $Q(t)$ that would improve the performance.  However, we can get a larger fraction transmitted if on the elevator for $\pi\leq t$(mod$2\pi)<2\pi$ a pusher shoves the passengers to the right so they exit the elevator more quickly.

Clearly, when the energy of incident particles
is less than the amplitude of the static barrier, the
only possible direction of fractional particle transport is left-to-right.
This is because all particles approaching from the right
are reflected, while some particles incident from the left
may hop onto the oscillating barrier and gain enough energy
from it to transmit over the static barrier. On the other hand,
suppose the energy of incident particles is higher than the
peak of the static barrier, and suppose that an equal number of particles
approaches from the left and from the right.  Then the only possible direction
of net fractional transport is the opposite direction, from right-to-left. In this
regime, all particles approaching from the right transmit
over both barriers, but particles incident from the left
may lose energy while riding the oscillating barrier down, and
can then be reflected from the static barrier.

The pumping mechanism of a diode is easily pictured by thinking
about rectangular barriers, but it also applies to smooth barriers
with smooth time dependence.  To keep the analysis simple, let us consider rectangular elevators with some smooth dependence on $t$.  Again particles approach from the left with fixed kinetic energy $K_i$, and uniform spatial density.  Let $t_{-b}$ be the time that a particle arrives at the point $x=-b$.  It is reflected if $K_i<Q(t_{-b})$;  otherwise it jumps onto the elevator and moves across it with constant kinetic energy
\begin{equation}
K_L = K_i - Q(t_{-b}).
\end{equation}
It reaches $x=0$ at time
\begin{equation}
t_0 = t_{-b} + \frac{b}{\sqrt{2K_L}} = t_{-b} + \frac{b}{\sqrt{2\left(K_i - Q(t_{-b})\right)}}\label{t0}
\end{equation}
when its total energy is
\begin{equation}
E_{0} = K_i +Q(t_0) - Q(t_{-b}).
\end{equation}
(Here the index $0$ does not mean ``initial,'' but rather ``when the particle arrives at $x=0$.'')  If $E_0<\hat{U}$, the particle is reflected by the right-hand barrier.  Otherwise it is transmitted, with kinetic energy
\begin{equation}
K_R = E_0 - \hat{U}.
\end{equation}

At $x=b$, its potential energy is converted to kinetic energy, and it escapes to the right with kinetic energy $K_f = E_0$.  Summarizing, for $0<t_{-b}<2\pi$ and initial kinetic energy $K_i$, we get transmission with final kinetic energy $K_f = K_i+Q(t_0) - Q(t_{-b})$ provided that i.)
$K_i>Q(t_{-b})$,  and ii.) $K_i +Q(t_0) - Q(t_{-b}) >\hat{U}$, where $t_0$ is given by Eq.~\eqref{t0}.

Each particle trajectory beginning at $x_{i_k}$ and ending near momentum $p_f = (2 m K_f)^{1/2}$  contributes a term to the classical probability density $P^C(p_f)$ , given by
\begin{align}
P^C(p_f)&=\sum\limits_{k}\left|\Psi\left(x_i(p_f),t_i=0\right)\right|^2\left|\frac{\partial p_f}{\partial x_{i}}\right|^{-1}_{x_i=x_{i_k}(p_f)}\nonumber\\
&=\sum\limits_{k}\left|\Psi\left(x_i(p_f),t_i=0\right)\right|^2|\tilde{J}_k(p_f)|^{-1},
\end{align}
where $\tilde{J}_k(p_f)$ is the Jacobian for the $k^{th}$ trajectory ending near $p_f$. Summing over all trajectories gives a smooth result which diverges at extrema of $\mathcal{P}_f(x_i)$.

The ``primitive'' semiclassical wave function in momentum space is obtained via the same method as in \cite{single}, and is similar to the methods in \cite{MasFed,delos863,schwaldelos,spellabunch,haggplus3,haggdelos,cdsanddelos1,cdsanddelos2,dudelos,dudelos2,dudelos3}. For each $p_f$ at the final time $t_f$, we sum semiclassical terms
\begin{align}
\tilde{\Psi}_k^{SC}\left(p_f,t_f\right)&= F\left(x_i(p_f,t_f)\right)|\tilde{J}_k\left(p_f,t_f\right)|^{-1/2}\nonumber\\
&\times\exp\left(i \tilde{\mathcal{S}}_k\left(p_f,t_f\right)/\hbar\right)\exp\left(-i \tilde{\mu}_k \pi/2\right),\label{scpsi}
\end{align}
where $\tilde{\mu}_k$ is the Maslov index for the $k^{th}$ branch of the function $p_f(x_f)_{t=t_f}$, and
\begin{equation}
\tilde{\mathcal{S}}_k\left(p_f,t_f\right) = -\int \left[x(t) \frac{dp(t)}{dt}\right]dt - \int E(t)dt
\end{equation}
is integrated over the classical path from initial to final time. The primitive semiclassical approximation in Eq.~\eqref{scpsi} applies only in classically-allowed regions, and it diverges at the boundaries of these regions. However, the divergences can be repaired and the function can be extended into classically-forbidden regions via the method in \cite{single}.

\begin{figure}[t]
\includegraphics*[width=\columnwidth]{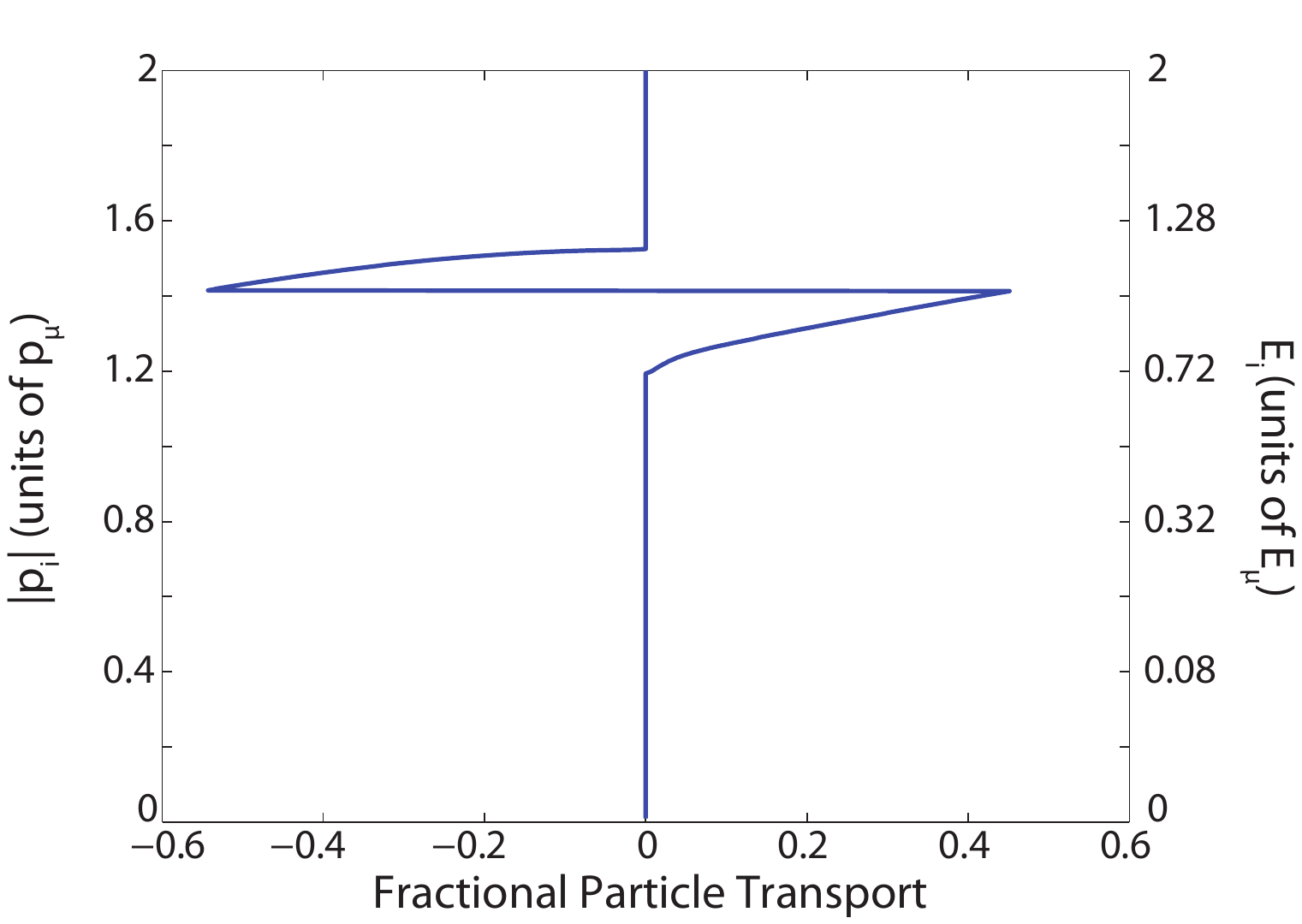}
\caption{(Color Online) (a) Fractional transport of particles, $C_P(|p_i|)$, for the diode described by Eq.~\eqref{1strec} with $\alpha=0.9$, $\omega=0.07$, $b=5$, and $\hat{U}=1$. The incident energy $E_i$ corresponding to each $|p_i|$ is shown on the right-hand axis. The fractional transport abruptly switches direction at $E_i=\hat{U}$.}\label{rect_1stcurr}
\end{figure}

For rectangular barriers, particle momentum changes only at the barrier edges (i.e., $x=0$ and $x=\pm b$), so for particles which transmit past both barriers,
\begin{align}
\tilde{\mathcal{S}}_k\left(p_f\right) &= -\left[b\Delta p_b - b \Delta p_{-b}\right] -K_i t_{-b}\nonumber\\
& -\int^{t_0}_{t_{-b}}\left[K_L+Q(t)\right]dt -E_0\left(t_f - t_0\right)
\end{align}
where
\begin{align}
\Delta p_{-b} &=\sqrt{2mK_L} - \sqrt{2mK_i}\\
\Delta p_b &=\sqrt{2mK_f} - \sqrt{2mK_R}.
\end{align}

We now examine a diode described by
\begin{align}
\begin{array}{cllrlll}
U_L(x,t)&=0.5\left[1+\alpha \sin\left(\omega t\right)\right]&,&  -b&<x&<&0\\
U_R(x,t)&= \hat{U}&,&   0&<x&<&b,\label{1strec}
\end{array}
\end{align}
with $\alpha=0.9$, $\omega=0.07$, $b=5$, $\hat{U}=1$. The left barrier oscillates between a minimum value of $U_L=0.05$ and a maximum value of $U_L=0.95$, while the right barrier is static with a height of $\hat{U}=1$.

Fig.~\ref{rect_1stcurr} shows fractional transport $C_P(|p_i|)$ for this diode. In this example, when incoming particles have $E_i=K_i<\hat{U}$, $C_P(|p_i|)=0$ below the energy at which particles incident from the left begin to gain enough energy from the oscillating barrier to transmit past the static barrier. When particles incident from the left begin to transmit, particles incident from the right are all reflected, $C_P(|p_i|)>0$, and there is left-to-right fractional transport. As the incident particle energy increases, fractional transport monotonically increases until $E_i>\hat{U}$, which is the threshold energy for particles approaching from the right to transmit past both barriers. At this point, fractional transport abruptly reverses direction to right-to-left ($C_P(|p_i|)<0$). As $E_i$ increases, $C_P(|p_i|)\rightarrow0$ as particles incident from both sides transmit past both barriers.

We now analyze the behavior of particles with an initial energy of $E_i = 0.99$ (initial momentum $p_i = \pm \sqrt{2E_i}$). Particles incident from the right do not have enough energy to transmit past the right barrier, and are reflected with final energy $E_f=0.99$ (final momentum $p_f=\sqrt{2E_f}$). The initial wave packet approaching from the left has an envelope shape given by Eq.~\eqref{rectenv} centered at $-x_c=-1500$ with $\beta=300$. Particles incident from the left all have enough energy to hop onto the left barrier; approximately $43.4\%$ gain enough energy while traversing the left barrier to transmit past the right barrier, while the others are reflected from the right barrier.

\begin{figure}[t]
\includegraphics*[width=\columnwidth]{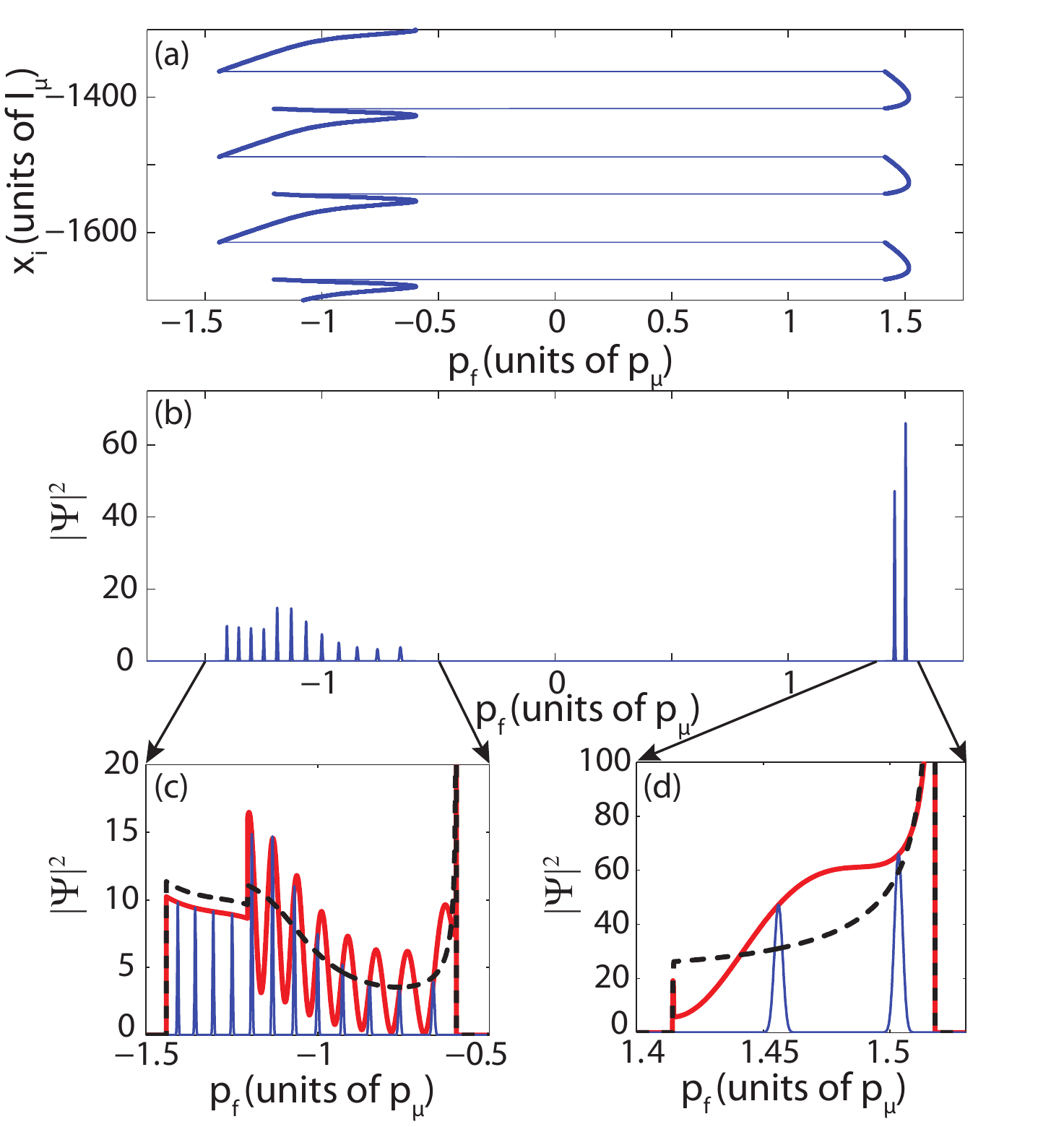}
\caption{(Color Online) (a) Initial position vs. final momentum for particles approaching the diode described by Eq.~\eqref{1strec} with $\alpha=0.9$, $\omega=0.07$, $b=5$, and $\hat{U}=1$ from the left with $p_i=\sqrt{2E_i}=\sqrt{1.98}$. (b) $P^{SC}_f(p_f)$, the absolute square of the primitive semiclassical wave function, for the particles in (a). (c) and (d) Expansions of (b). The dashed curve is the classical probability density, $P^C(p_f)$, for reflection or transmission with final momentum near $p_f$. The smoothly-varying solid curve (red online) is a single-cycle primitive semiclassical probability, $P^{SC}_s(p_f)$ \cite{single}. The sharp peaks are the full primitive semiclassical probability summed over all cycles. These occur at momenta corresponding to Floquet energies.}\label{rectrect}
\end{figure}

Fig.~\ref{rectrect} shows classical and semiclassical results for particles approaching from the left. Fig~\ref{rectrect}(a) shows initial position as a function of final momentum, $x_i(p_f)$. Only a small portion of initial positions are shown. Since the wave packet is wide in position space ($\Delta x_i > > L$), there is a periodic relationship between final momentum and initial position. Because the potential is not smooth, $x_i(p_f)$ is discontinuous between transmitted and reflected portions. Each branch of the function $x_i(p_f)$ contributes a term $\tilde{\Psi}_n^{SC}\left(p_f\right)$ to the primitive semiclassical wave function, given by Eq.~\eqref{scpsi}. The complete primitive wave function $\Psi^{SC}_f(p_f)$ is obtained by summing Eq.~\eqref{scpsi} over all branches of $x_i(p_f)$. Fig.~\ref{rectrect}(b) shows $P^{SC}_f(p_f)=|\Psi^{SC}_f(p_f)|^2$, the final primitive semiclassical probability density. Fig~\ref{rectrect}(a) shows that many trajectories end with any given $p_f$ inside the classically-allowed regions. The sharp peaks in $P^{SC}_f(p_f)$ arise from interference among all trajectories ending with any given $p_f$. This calculation has not been extended into the classically-forbidden regions, so all peaks lie within the classically-allowed regions for both the transmitted and reflected portions.

Since $P^{SC}_f(p_f)$ includes interference from a great number of trajectories, it is useful to differentiate between two distinct types of interference: i) interference from within a single cycle of $x_i(p_f)$ (intracycle interference), and ii) interference among all cycles (intercycle interference). To view intracycle interference, we choose an arbitrary $x_i(p_f)$ and sum the corresponding $\tilde{\Psi}_n^{SC}\left(p_f\right)$ terms from within one cycle of the chosen $x_i(p_f)$ to obtain $\Psi^{SC}_s(p_f)$.

Figures~\ref{rectrect}(c) and (d) show the classical probability density $P^C(p_f)$ (dashed curve). Note that the scales are different in Figs.~\ref{rectrect}(c) and (d). We see that whereas classical theory gives a slowly-varying probability density $P^C(p_f)$, the primitive semiclassical single-cycle probability density $P^{SC}_s=|\Psi^{SC}_s(p_f)|^2$ (thick solid curve, red online) is oscillatory. The oscillations arise from interference among trajectories in the cycle that end with the same final momentum. The discontinuities seen in $P^{SC}_s(p_f)$ and $P^C(p_f)$ in Fig.~\ref{rectrect}(c) at $p_f\approx -1.2$ are due to the behavior of the branches in $x_i(p_f)$ in Fig.~\ref{rectrect}(a): for $p_f \gtrsim -1.2$ within the classically-allowed final momentum region of the reflected segments, there are two interfering branches per cycle, but for $p_f \lesssim -1.2$, there is only one branch per cycle.

Summing  $\Psi^{SC}_s(p_f)$ over all cycles yields the full primitive wavefunction $\Psi^{SC}_f(p_f)$, the square of which is $P^{SC}_f(p_f)$, the sharply-peaked function in Figs.~\ref{rectrect}(b), (c), and (d). This function has peaks at energies $E_n=K_i +n\hbar \omega$, consistent with Floquet theory. In Figs.~\ref{rectrect}(c) and (d), $P^C(p_f)$ and $P^{SC}_s(p_f)$ are scaled (multiplied by the same constant). When plotted in this fashion, one can see that the relative heights of the peaks in $P^{SC}_f(p_f)$ closely align with $P^{SC}_s(p_f)$, i.e., the relative heights of the Floquet peaks are governed by the single-cycle probability. This occurs for any arbitrary $x_i(p_f)$ chosen as the beginning of a cycle; while different choices yield different $P^{SC}_s(p_f)$, they all intersect at the locations of the Floquest peaks.

\subsection{Quantum Suppression of Classical Transmission}
Another interesting phenomenon arising from a similar elevator system is the quantum suppression of classical transmission. It may happen that the classical transmission probability is large, but the range of transmitted momenta is small -- so small that no Floquet peaks lie in the classically-allowed range. Then quantum interference (we might better say semiclassical interference) among trajectories from different cycles prevents transmission that is classically allowed. In such a case, a narrow initial wave packet (in $x_i$) may allow transmission both classically and quantum-mechanically, not because it is broad in momentum space, but because it interacts with the barrier for only one (or a few) cycles.

\begin{figure}[t]
\includegraphics*[width=\columnwidth]{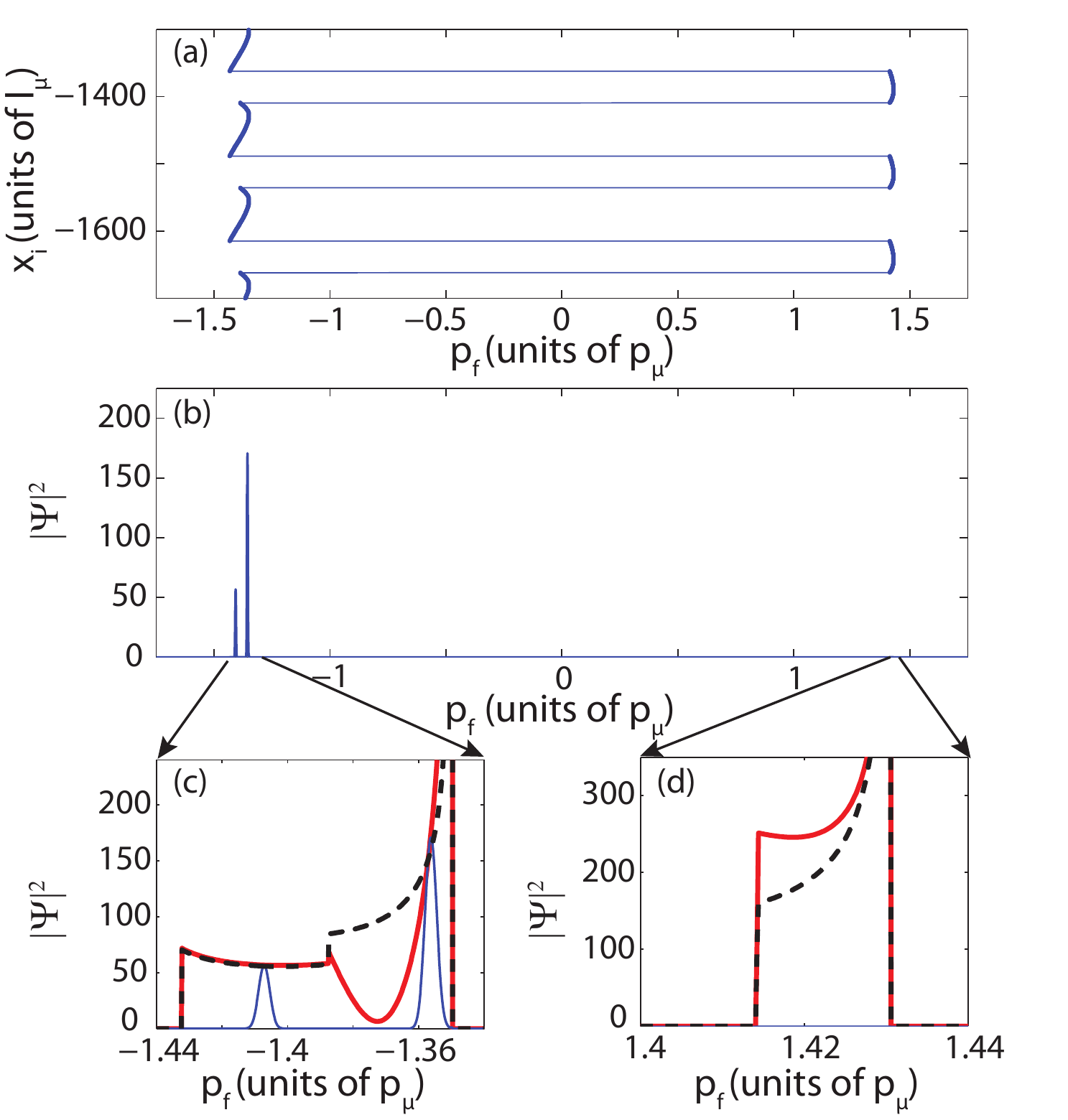}
\caption{(Color Online) Quantum interference suppresses classical transmission for the diode described by Eq.~\eqref{rect2supp} with $\alpha=1-(.88/.92)\approx 0.0435$, $\omega=0.07$, $b=5$, and $\hat{U}=1$. All curves are as described in Fig.~\ref{rectrect}.}\label{rectanti}
\end{figure}

These phenomena occur for a diode described by

\begin{align}
\begin{array}{cllrlll}
U_L(x,t)&=0.92\left[1+\alpha \sin\left(\omega t\right)\right]&,&  -b&<x&<&0\\
U_R(x,t)&= \hat{U}&,&   0&<x&<&b,\label{rect2supp}
\end{array}
\end{align}
with $\alpha=1-(.88/.92)\approx 0.0435$, $\omega=0.07$, $b=5$, and $\hat{U}=1$. In this example, the left barrier oscillates between a minimum of $0.88$ and a maximum of $0.96$, and  particles approach the barriers from both sides with initial energy $E_i=0.99$ (initial momentum $p_i =\pm \sqrt{2E_i}$). Particles approaching from the right do not have enough energy to hop onto the right barrier, and are reflected with final energy $E_f=0.99$ (final momentum $p_f=\sqrt{2E_f}$). Particles approaching from the left all have enough energy to hop onto the left barrier, and classically, more than one third (approximately $37.3\%$) of these particles gain enough energy to transmit past the right barrier (see Fig.~\ref{rectanti}). These transmitted particles all end with $p_f$ inside a very small range.

For the semiclassical calculation, we took an envelope given by Eq.~\eqref{rectenv} with $-x_c=-1500$ and $\beta=300$. Fig.~\ref{rectanti}(b) shows the primitive semiclassical final momentum probability $P^{SC}_f(p_f)$ in the classically-allowed regions. In contrast to the classical result, we see no visible transmission. Figures~\ref{rectanti}(c) and (d) show the classical transmission probability and the single-cycle and final primitive semiclassical probabilities (similar to Fig.~\ref{rectrect}). The single-cycle primitive semiclassical calculation gives an even larger total transmission than the classical result, but the final semiclassical result is essentially zero.

The explanation is that Floquet peaks occur at energies $E_n=K_i+n\hbar \omega$, and the corresponding momenta for $n=(-1,0,1)$ are $p_n \approx 1.36$, $1.41$, and $1.46$. None of these momenta lie inside the classically-allowed region of transmission. Therefore, when summing interference from all cycles, this interference is destructive across the entire range of transmitted momentum, and at this level of approximation, there is no transmission. (A uniform semiclassical approximation would extend into classically-forbidden regions, but the decay of the wave function in these regions combined with the Floquet ``comb'' would yield small peaks, comparable to those seen outside the classically-allowed regions in Fig.~\ref{gauss2osc} and in Ref. \cite{single}).

In this example, quantum interference suppresses the classical probability density for transmitted particles.

\subsection{Gaussian Barriers}

\begin{figure}[t]
\includegraphics*[width=\columnwidth]{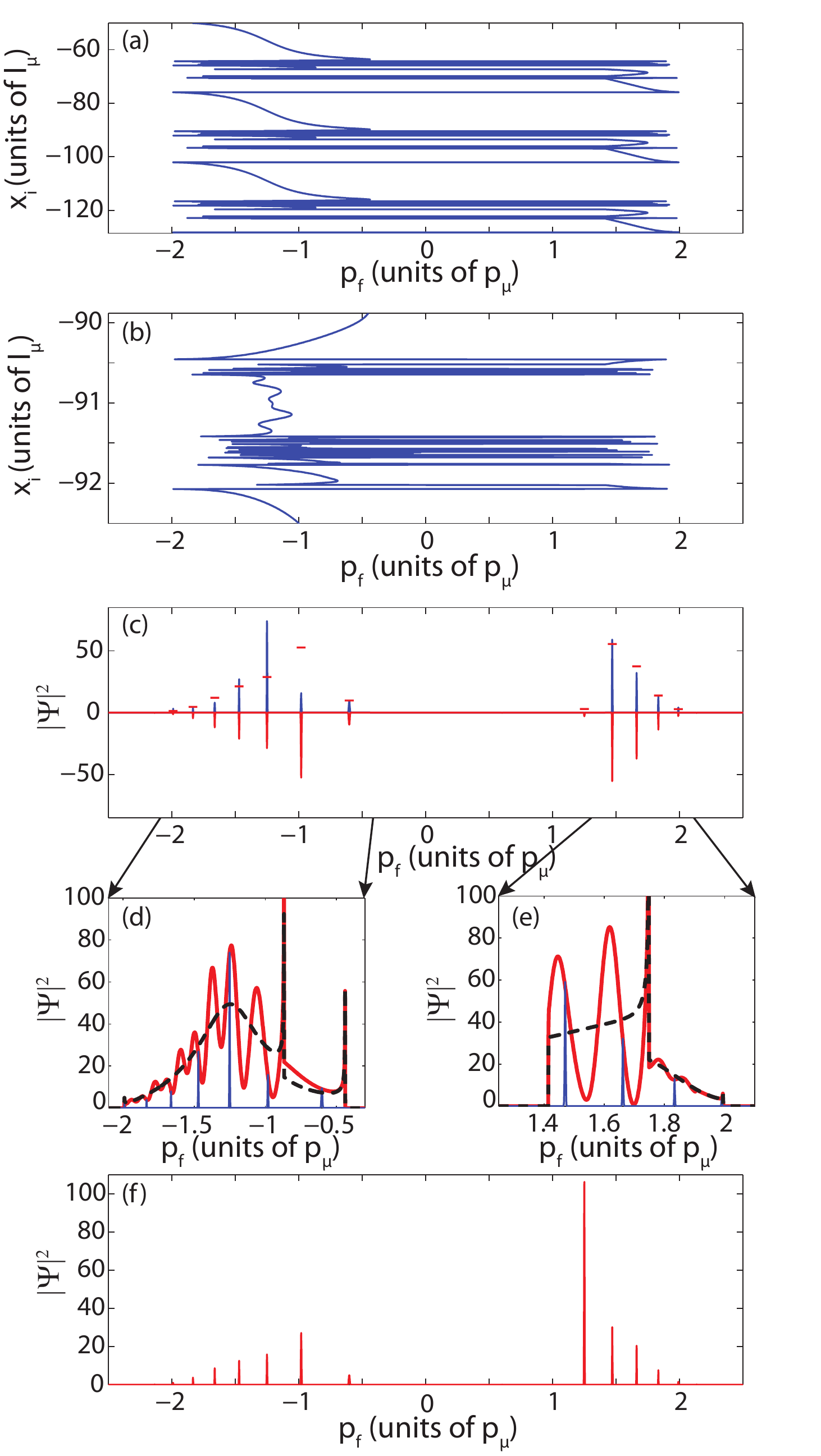}
\caption{(Color online) Dynamics for Gaussian diode. (a) Three cycles of $x_i(p_f)$ for the packet approaching a Gaussian diode from the left. (b) Zoom of (a), showing the complexity of chaotic trajectories. (c) Quantum (downward, red)  and primitive semiclassical (upward, blue) final momentum probabilities. The semiclassical calculation only includes the contributions of slowly-varying branches of $x_i(p_f)$. (d) and (e) $P^C(p_f)$ (dashed curve), $P^{SC}_s(p_f)$ (oscillatory curve, red), and $P^{SC}_f(p_f)$ (sharply-peaked curve, blue). (f) Quantum-mechanical final momentum probability for incoherent packets approaching the barriers from both sides.}\label{gaussrects}
\end{figure}

A more realistic type of diode is one which has Gaussian barriers described by Eqs.~\eqref{gaussleft} and \eqref{gaussright} with $\alpha_R=0$. We examine one such case with barriers described by $\hat{U}_R=\hat{U}_L=1$, $\alpha_L=1$, $\omega=0.30$, $\sigma=2.5/2\sqrt{2\ln{2}}$ (full width at half maximum of $2.5$), and $\hat{x}=3.75$. The right barrier has static height $\hat{U}_R=1$ and the left barrier oscillates between zero and twice the height of the static barrier. Fig.~\ref{gaussrects} shows classical, semiclassical, and quantum calculations for particles incident on this diode from both directions with $|p_i|=\pm1.25$ ($E_i\approx .78$). Particles incident from the right with this inital energy are all classically reflected, but approximately $30.3\%$ of particles incident from the left transmit, and there is left-to-right fractional transport of particles. Fig.~\ref{gaussrects}(a) shows classical $x_i(p_f)$ for particles incident from the left. Classical trajectories are chaotic, as some particles are reflected from the left oscillating barrier, others directly transmit past both barriers, and others are temporarily trapped between the barriers before finally reflecting or transmitting. Fig.~\ref{gaussrects}(a) shows three periods of the function $x_i(p_f)$, and Fig.~\ref{gaussrects}(b) shows a zoom consisting of 10\% of a period (30X magnification of (a)). Extreme dependence on initial position is apparent, and there are a large number of trajectories ending with any classically-allowed $p_f$.

Fig.~\ref{gaussrects}(c) shows quantum-mechanical (plotted downward, red online) and primitive semiclassical (plotted upward, blue online) final momentum probabilities for the packet incident from the left. The initial packet is described by Eq.~\eqref{rectenv} with $\beta=300$ and $-x_c=-1250$. The primitive semiclassical probability $P^{SC}_f(p_f)$ only includes contributions from slowly-varying branches of the function $x_i(p_f)$ (i.e., regions of chaotic scattering are omitted). This rough approximation agrees reasonably well with the quantum probability $P^Q_f(p_f)$ except for peaks located at $p_f=-1.25$ and $p_f\approx-0.98$. Figs.~\ref{gaussrects}(d) and (e) show zooms of the primitive semiclassical approximation, along with the single-cycle momentum probability $P^{SC}_s(p_f)$ (thick oscillatory curves, red online) and classical momentum probability $P^C_f(p_f)$ (dashed curve). As before, the single-cycle probability governs the relative heights of the Floquet peaks seen in the full primitive semiclassical probability.

Fig.~\ref{gaussrects}(f) shows the quantum-mechanical final momentum probability for particles approaching the barriers from both sides with $|p_i|=\pm1.25$. The largest probability is at $p_f=1.25$, which is primarily caused by the reflection of particles incident from the right. Particles incident from the left that are transmitted contribute only a small amount to this momentum state (see Fig.~\ref{gaussrects}(c)); these transmitted particles have a much higher probability of ending with $p_f=\sqrt{2(E_i+n\hbar\omega)}$ with $n=1,2,3$. The total probability in the quantum calculation for $p_f>0$ is approximately $65.9\%$. The classical fractional transport $C_P(|p_i|=1.25)\approx 0.303$ corresponds to approximately $65.1\%$ of particles ending with $p_f>0$, showing good agreement between classical and quantum theories.


\section{\label{nopump}Symmetric Pumps: A General Theorem}

In the remainder of this paper, we consider pumps that are ``symmetric'' in the sense that
$\hat{U}_L=\hat{U}_R=\hat{U}$, $\alpha_L=\alpha_R=\alpha$, and $\sigma_L=\sigma_R=\sigma$,
so the barriers are identical, but not in
phase with each other.  Intuitively one might have guessed the following behavior.  Suppose that we consider the case of the classic turnstile pump for which
$\phi=-\pi/2$, so the barrier on the right oscillates a quarter-cycle behind the one on the
left. Then the two barriers together imitate a rightward-moving wave, $\sin(kx-\omega t)$.
We might then expect that the system would preferentially pump particles from left to right.

Nothing of the sort happens, however.  Classically, if particles begin with a
distribution that is uniform in {\it both} momentum and position ({\it i.e.} the distribution
includes all initial energies $E_i$ and is independent of $E_i$)
then for every particle going from left to
right, another goes from right to left -- there is no net pumping at all.

This symmetry theorem can be violated if the initial distribution of particles is not uniform in phase space.
For example, if the phase space distribution is constant only up to some maximum initial energy, then
some net pumping is possible.  More important, if particles begin from both sides with the
same fixed energy, then there can
be a net flow in one direction or the other.  The amount and direction of this flow
depends on that energy, so the apparent natural direction of the pump is an illusion.

\begin{figure}
\includegraphics[width=.40 \textwidth]{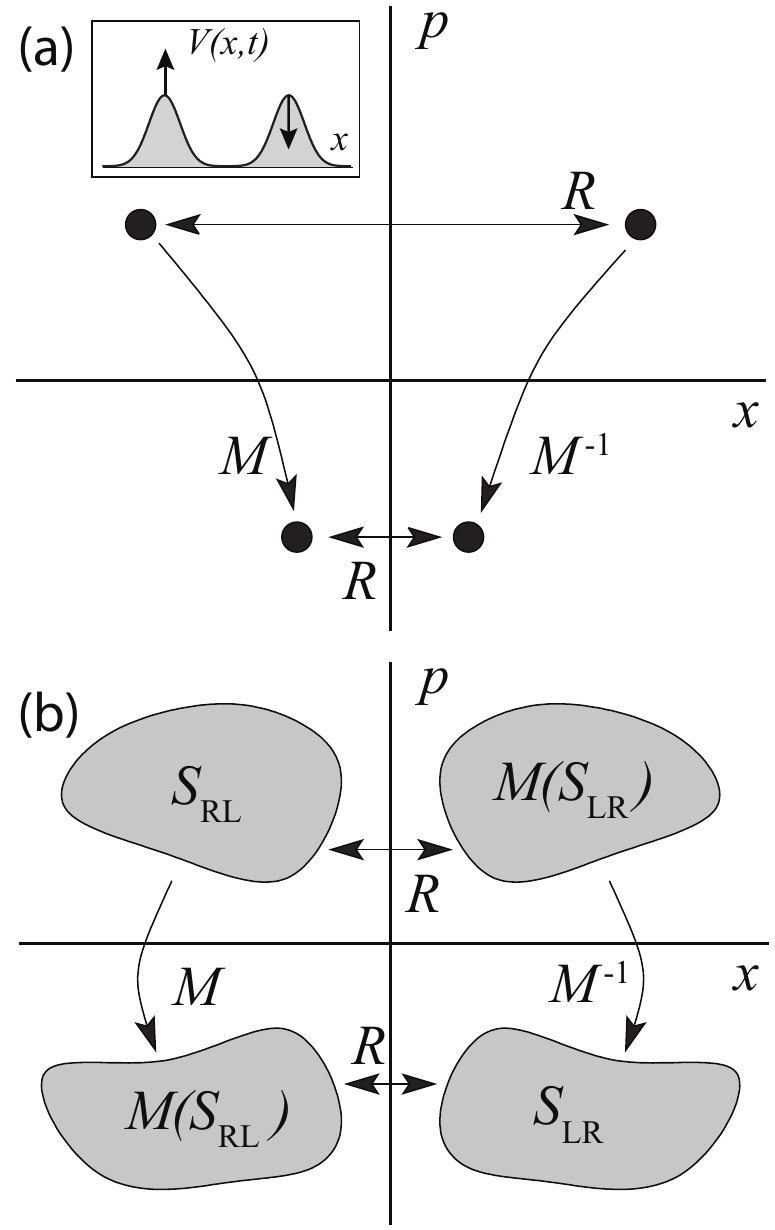}
                \caption{\label{fig:NoPumpingThm} (a) An illustration
                  of Eq.~(\ref{r1}); The phase of the sampling is
                  chosen so that both barriers have the same height
                  with the left barrier moving up and the right moving
                  down (inset).  (b) Illustration of the sets $S_{RL}$
                  and $S_{LR}$ and the relation $S_{LR} =
                  RM(S_{RL})$.  The double arrows reflect the fact $R
                  = R^{-1}$.}
\end{figure}

The critical step in proving this no-pumping result is
choosing a reference phase of the oscillations, and then
using a surface of section at integer number of cycles
from this reference phase.  We choose the reference phase
to be when the two oscillating
barriers have equal height, and
the left barrier is going up and the right barrier is
going down (see Fig. \ref{fig:NoPumpingThm}(a)(inset)).  If we take such a point in any
cycle to be $t=0$, then for any $\phi$ and all $t$ and $x$ we have
\begin{eqnarray}
V(-x,-t)=V(x,t).
\end{eqnarray}
Let $M:(x,p) \mapsto (x',p')$ be the map that evolves a point $(x,p)$
forward one pumping period to $(x',p')$, and let $R=R^{-1}$ be the operator that reflects the position through the origin: $R(x,p)=(-x,p)$. A trajectory on the left of
the pump moving right sees the closest barrier going
up, whereas a trajectory on the right moving left sees its
closest barrier going down. A mirror image thus converts
the upward moving barrier to downward and vice-versa,
i.e. it reverses the time-dependence of the barriers, so the particle follows the time-reversed trajectory. Consequently, $M^{-1}(-x,p)$ will be a mirror image of $M(x,p)$. More formally,
\begin{equation}
M^{-1} = R M R.
\label{r1}
\end{equation}
This relation
is demonstrated in Fig.~\ref{fig:NoPumpingThm}(a).

Now define $S_{RL}$ to
be the set of points moving to the right that are reflected after one
pumping period, i.e.
\begin{equation}
S_{RL} = \{ (x, p) | p>0, p'<0 \mbox{ where } (x',p') = M(x,p) \}.
\end{equation}
We define $S_{LR}$ similarly
\begin{equation}
S_{LR} = \{ (x, p) | p<0, p'>0 \mbox{ where } (x',p') = M(x,p) \}.
\label{r2}
\end{equation}
\emph{Assuming a uniform initial distribution in phase space}, the
single-period net flux $F$ of rightward to leftward moving
trajectories is thus
\begin{equation}
F = \mbox{area}(S_{RL}) - \mbox{area}(S_{LR}).
\end{equation}

We now show $S_{LR} = RM(S_{RL})$.  (See
Fig.~\ref{fig:NoPumpingThm}b.)  Let $(x,p) \in S_{RL}$ be arbitrary.
We then have $p>0$ and $p'<0$ where $(x',p') = M(x,p)$.  The point $(x'',p'')
= RM(x,p)$ is then an arbitrary point of $RM(S_{RL})$; note $p''=p'$.
Now,
\begin{equation}
M(x'',p'') = MRM(x,p) = R(x,p) = (-x,p),
\end{equation}
where the second equality follows from Eq.~(\ref{r1}).  Since
$p''=p'<0$ and $p>0$, we find $(x'',p'') \in S_{LR}$.  Hence,
$RM(S_{RL}) \subset S_{LR}$.  The reverse inclusion follows similarly.

Since $S_{LR} = RM(S_{RL})$ and $R$ and $M$ both preserve phase-space
area, $\mbox{area}(S_{RL})$ = $\mbox{area}(S_{LR})$.  Hence $F = 0$,
i.e. there is no net flux pumped across the barrier. All of our numerical simulations of symmetric pumps have confirmed this theorem.

\section{\label{symm}Symmetric Rectangular Barriers}

\subsection{\label{symmnospace}No space between the barriers}

We now consider symmetric turnstile pumps in which both barriers oscillate smoothly in time, with rectangular potential barriers described by Eqs.~\eqref{recleft} and \eqref{recright}. We first examine the simplest pump of this type, which has no space between the barriers.  In this case, any incident particle can either be reflected by the first barrier, hop onto the first barrier and be reflected from the second barrier, or transmit over both barriers.  If the particle has enough energy to transmit over one or both barriers, it can gain or lose energy during the time it spends on top of the barrier(s).

As in Section~\ref{RecElev}, since the barriers are rectangular, particles only experience acceleration at the boundaries of barriers, and have constant momentum everywhere else.  A particle beginning to the left of the barriers is launched with momentum $p_i>0$ and arrives at the leftmost edge of $U_L$ at time $t_{-b}$, at which time the height of the left barrier is $U_L(t_{-b})$, and the total energy of the particle is $E_i=p_i^2/2$.  If $E_i\le U_L(t_{-b})$, the particle is reflected from the first barrier with final momentum $p_f=-p_i$.  Otherwise, the particle is transmitted over the first barrier with momentum

\begin{eqnarray}
p_{b_1}=\sqrt{2(E_i-U_L(t_{-b}))}.  \label{pa}
\end{eqnarray}

The time at which the particle reaches the opposite edge of the first barrier (and therefore the first edge of the second barrier) is

\begin{eqnarray}
t_0=\frac{2\sigma}{p_{b_1}}+t_{-b}.\label{ta}
\end{eqnarray}
The corresponding $U_L(t_0)$ and $U_R(t_0)$ are given by Eqs.~\eqref{recleft} and~\eqref{recright}, respectively, and $E(t_0)=U_L(t_0)+p_{b_1}^2/2$.  If $E(t_0)\le U_R(t_0)$, the particle is reflected from the second barrier with $p=-p_{b_1}$ and spends another time interval $2\sigma/p_{b_1}$ going back over the first barrier, after which it falls off  onto the left-hand side of the pump with final momentum

\begin{eqnarray}
p_f=-\sqrt{2\left[U_L\left(t_0+\frac{2\sigma}{p_{b_1}}\right)+\frac{p_{b_1}^2}{2}\right]}.
\end{eqnarray}

If $E(t_0)>U_R(t_0)$, the particle is transmitted over the second barrier with momentum
\begin{eqnarray}
p_{b_2}=\sqrt{2(E(t_0)-U_R(t_0))}.
\end{eqnarray}
The time at which the particle falls off the second barrier is
\begin{eqnarray}
t_b=\frac{2\sigma}{p_{b_2}}+t_0 ,
\end{eqnarray}
at which time the height of the right-hand barrier is $U_R(t_b)$, and the final momentum is
\begin{eqnarray}
p_f=\sqrt{2\left(U_R(t_b)+\frac{p_{b_2}^2}{2}\right)}.\label{endsec}
\end{eqnarray}

A similar algorithm is followed for particles beginning on the right of the pump with negative initial momentum.  There is never more than one reflection of a particle. We calculate all particle trajectories using Eqs.~\eqref{pa}-\eqref{endsec} to obtain each particle's final momentum $p_f$.

We examine the net particle fractional transport for mirrored sets of particle packets approaching the barriers from opposite directions with $\pm p_i$. Classical computations shown in the remaining sections are done as follows. For a selected set of barrier parameters, we first choose a range of initial particle energies $\Delta E_i$. Each $E_i$ in this range has two corresponding momenta, $\pm p_i$. For each $|p_i|$, we construct two incoming packets of particles: one starts to the left of the barriers with $p_i = +|p_i|$, and the other starts to the right of the barriers, with $p_i = -|p_i|$. The width in $x_i$ of each packet is $\Delta x_i = |p_i|T = |p_i|2\pi/\omega$ (recall $m=1$). We start all trajectories at $t_i=0$. The edge of each packet which is closest to the barriers is placed a distance $d=|p_i|2\pi/\omega$ away from the outer edge of the first barrier, which ensures that the first particle of each packet reaches the outer edge of the first barrier at $t=2\pi/\omega$.

For these initial conditions, we define the time of arrival at the barrier as $\theta=t-2\pi/\omega$; with this definition, the particle in each packet which starts closest to the barriers arrives at the outer edge of the first barrier at $\theta=0$. Our choice of packet width ensures that the last particle in each packet to arrive at the outer edge of the first barrier arrives at $\theta=2\pi/\omega$, which represents one full cycle of the barriers. Referring to Eq.~\eqref{recleft}, a particle that arrives at the left-hand edge of the left barrier at $\theta=0$ or $\theta=2\pi/\omega$ encounters the barrier at its maximum, and one that arrives at $\theta=\pi/\omega$ encounters that barrier at its minimum.

\begin{figure}[t]
\includegraphics*[width=\columnwidth]{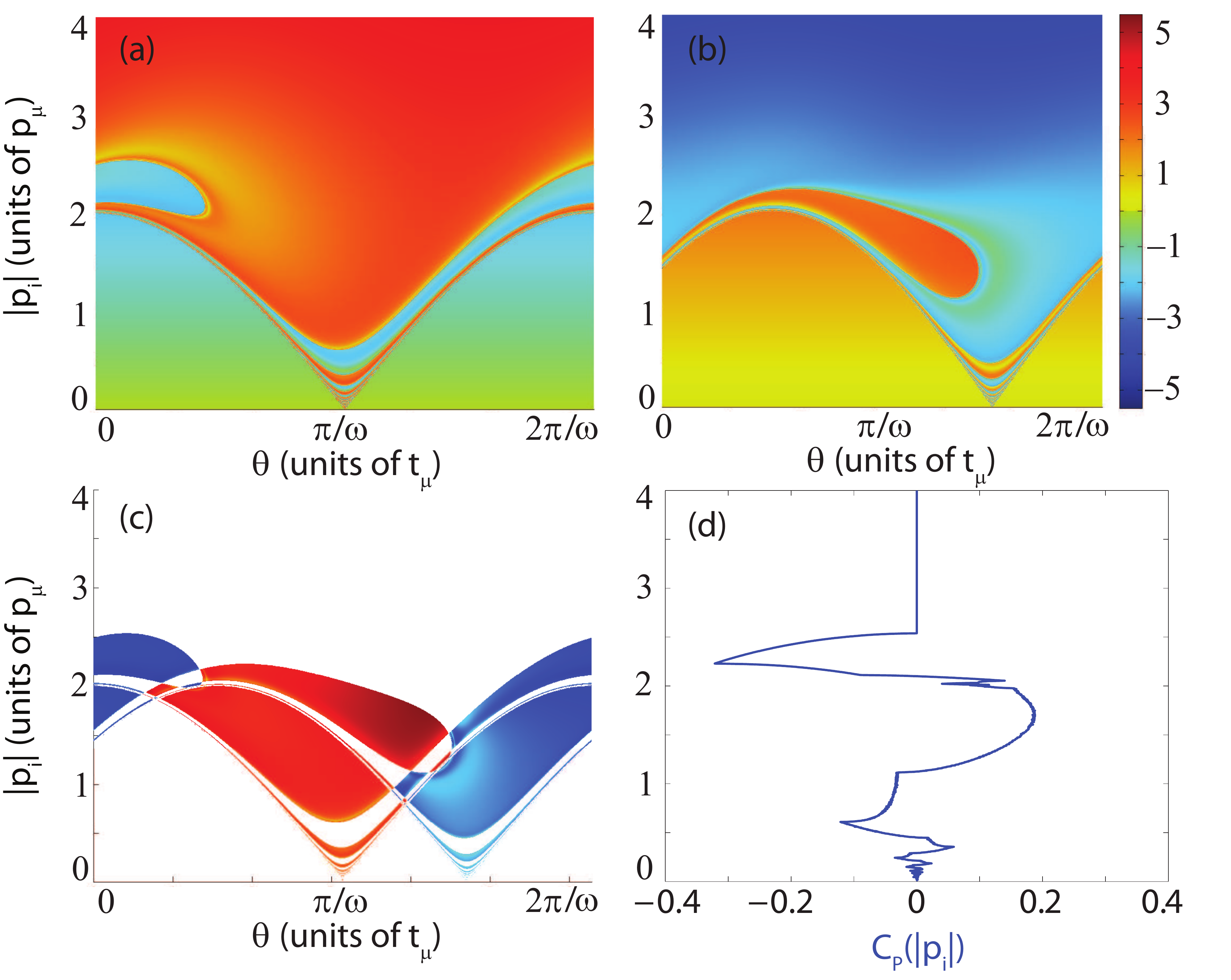}
\caption{(Color online) (a) and (b) $p_f(|p_i|,\theta)$ for particles incident on rectangular barriers described by $\hat{U}=1$,  $\omega=1$, $\phi=3\pi/2$,  $\hat{x}=\sigma=1.25$ and $\alpha=1$. (a) represents particles approaching from the left, and (b) represents particles approaching from the right. (c) Sum of $p_f(|p_i|,\theta)$ for particles approaching the barriers from both directions with $p_i=\pm |p_i|$. Red indicates both particles scatter to the right, and blue indicates both particles scatter to the left. (d) Fractional transport, $C_P(|p_i|)$.}\label{rbnospace}
\end{figure}

The numerical results for barriers with $\hat{U}=1$,  $\omega=1$, $\phi=3\pi/2$,  $\hat{x}=\sigma=1.25$ and $\alpha=1$ are shown in Fig.~\ref{rbnospace}. Figure~\ref{rbnospace}(a) represents particles approaching the barriers from the left, and ~\ref{rbnospace}(b) represents particles approaching from the right. In both plots, individual particles are represented by their initial momenta $|p_i|$ and the time $\theta$ at which they arrive at the outer edge of the first barrier. The colors (online) in both plots correspond to the final momentum $p_f=p_f(|p_i|,\theta)$ of each particle. Blue (online) represents particles which scatter to the left ($p_f(|p_i|,\theta)<0$) of the barriers, and red (online) corresponds to particles which scatter to the right of the barriers $(p_f(|p_i|,\theta)>0)$. The intensity of the color corresponds to the magnitude of the particle's final momentum, as seen in the colorbar.

The lowermost blue region in Fig.~\ref{rbnospace}(a) represents particles approaching from the left that are initially reflected from the left barrier, and the lowermost yellow region in Fig.~\ref{rbnospace}(b) represents particles approaching from the right that are directly reflected by the right barrier. In both cases, if there were no other barrier, then the region above this lowermost boundary would be entirely of the opposite color, as all particles not initially reflected would be transmitted. It follows that all of the striping effects just above this boundary are due to the presence of the second barrier.  Just above this boundary, a particle has just enough energy to hop onto the first barrier it encounters. Consequently, its momentum $p_{b_1}$ on the first barrier is small, it moves across the barrier slowly, and the barrier may oscillate many times while the particle is on it. In the limit that $p_{b_1} \rightarrow 0$, an infinite number of oscillations occurs while the particle is on the barrier.  Hence, there is an infinite number of stripes converging from above upon the boundary.

Fig.~\ref{rbnospace}(c) sums $p_f(|p_i|,\theta)$ for both particles which arrive at the barriers at the same time $\theta$ and $|p_i|$, but which arrive from opposite directions. Red (online) represents cases in which both particles scatter to the right of the barriers ($p_f(|p_i|,\theta)>0$ for both particles). Blue (online) represents cases in which both particles scatter to the left of the barriers ($p_f(|p_i|,\theta)<0$ for both particles). If the particles scatter to opposite sides of the barrier, e.g. if both are reflected or transmitted, no color is plotted. The intensity of the color corresponds to the magnitude of the sum of $p_f(|p_i|,\theta)$ for both particles.

Fig.~\ref{rbnospace}(d) shows fractional particle transport $C_P(|p_i|)$ (see Eq.~\eqref{partcurr}). This function is considered over the entire range of $\theta$ for each $|p_i|$, i.e, $C_P(|p_i|)$ accounts for all particles at a given $|p_i|$. When $C_P(|p_i|)$ is averaged over all $|p_i|$, the symmetry theorem tells us that there is no net particle transport. However, there is transport (in either direction) within finite ranges of $|p_i|$. Fractional particle transport at a given $|p_i|$ can be understood by comparing Figs.~\ref{rbnospace}(c) and (d). $C_P(|p_i|) <0$ in the range $2.2 \lesssim |p_i| \lesssim 2.5$, indicating net particle flow to the left of the barriers. Examining Fig.~\ref{rbnospace}(c), we see that only one colored lobe extends into this $|p_i|$ range. Its color (blue online) indicates $(|p_i|,\theta)$ values for which both particles have $p_f(|p_i|,\theta)<0$, meaning that both particles scatter to the left of the barriers. Since no red (online) lobes extend into this $|p_i|$ range, there are no $(|p_i|,\theta)$ values for which both particles scatter to the right. Therefore, for any $(|p_i|,\theta)$ in this range, both particles can either scatter to the left of the barriers, or scatter to opposite sides, causing net particle transport to the left.

\begin{figure}[t]
\includegraphics*[width=\columnwidth]{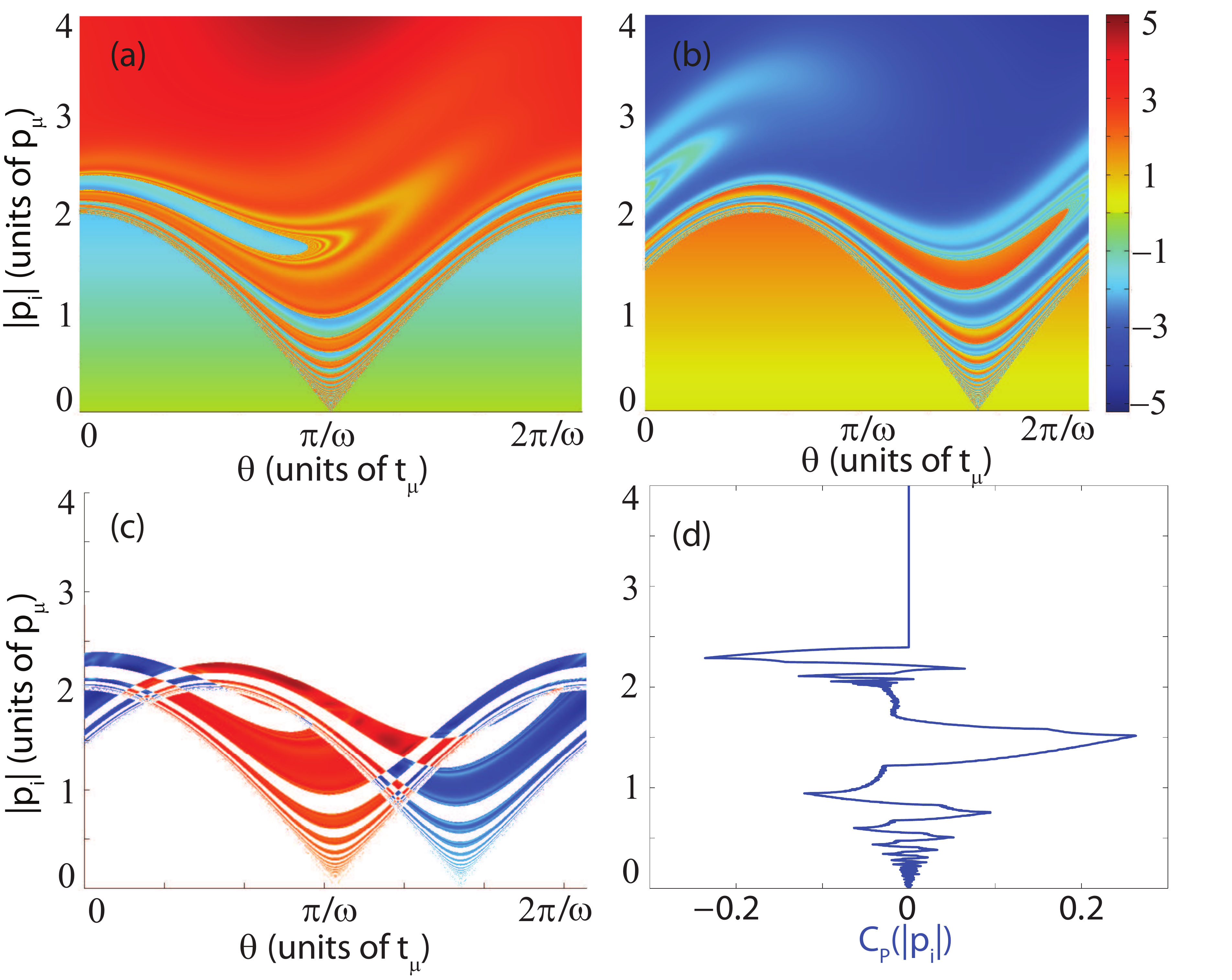}
\caption{Same as Fig.~\ref{rbnospace}, but with barriers four times as wide as those in Fig.~\ref{rbnospace} ($\sigma$ and $\hat{x}$ have been increased from $1.25$ to $5$). Increasing barrier width causes thinner ribbons of transmission and reflection for particles incident on the barriers from both sides.}\label{wider}
\end{figure}

Figure~\ref{wider} illustrates the effects of increasing the barrier widths. In this calculation, all parameters are the same as those in Fig.~\ref{rbnospace} ($\hat{U}=1$, $\alpha=1$, $\omega=1$, and $\phi=3\pi/2$), except for $\sigma$ and $\hat{x}$, which have been increased from $1.25$ to $5$. One can see that the ribbons of transmission and reflection span a more narrow $\Delta |p_i|$ range at a given $\theta$, and that the widths ($\Delta |p_i|$) of the ribbons at a given $\theta$ do not decay as rapidly as in the previous case as $p_i\rightarrow 0$. Comparing Figs.~\ref{wider}(c) and (d) to Figs.~\ref{rbnospace}(c) and (d), we see that in this case the net particle transport fluctuates more rapidly with $|p_i|$. However, the magnitude of $C_P(|p_i|)$ within these smaller $\Delta |p_i|$ regions can be just as large (or larger) as in the case of narrow barriers.

The change in transmission and reflection ribbons for wide barriers can be understood by examining the condition for particles to transmit. Let us examine particles which approach from the left with $p_i>0$ and arrive at the left barrier $\theta=\pi/\omega$, when the height of that barrier is zero. All particles arriving at $\theta=\pi/\omega$ hop onto the left barrier and traverse it with momentum $p_{b_1}=p_i$. The condition for them to transmit over the right barrier is
\begin{align}
f\left(p_i,t_0\right)=p_i^2/2+U_L(t_0)-U_R(t_0)>0.\label{feqlong}
\end{align}
When $f\left(p_i\right)>0$, particles transmit over the right barrier, and when $f\left(p_i\right)<0$, particles reflect from the right barrier. The zeroes of $f\left(p_i\right)$ thus mark the boundaries between transmission and reflection ribbons.

We illustrate this for the barrier parameters from the preceding two examples ($\hat{U}=\alpha=\omega=1$, $\phi=3\pi/2$). For these barrier parameters, Eq.~\eqref{feqlong} reduces to
\begin{align}
f\left(p_i\right)=\frac{p_i^2}{2}-2\cos\left(\frac{\pi}{4}+\frac{2\sigma}{p_i}\right)\sin\left(\frac{3\pi}{4}\right).\label{fzero}
\end{align}
Eq.~\eqref{fzero} shows that $f\left(p_i\right)$ oscillates about $p_i^2/2$, and the ratio $2\sigma/p_i$ governs its oscillation frequency. As $p_i\rightarrow 0$, $f\left(p_i\right)$ passes through zero an infinite number of times, resulting in an infinite number of transmission and reflection ribbons  for any $\sigma$. Higher $\sigma$ values (wider barriers) cause $f\left(p_i\right)$ to oscillate more rapidly as  $p_i\rightarrow0$. The maximum amplitude of oscillation is $2\sin(3\pi/4)=\sqrt{2}$; thus, for $p_i^2/2>\sqrt{2}$, i.e. $p_i>2^{3/4}\approx 1.68$, all particles will transmit, no matter the width of the barriers.

\begin{figure}[t]
\includegraphics*[width=\columnwidth]{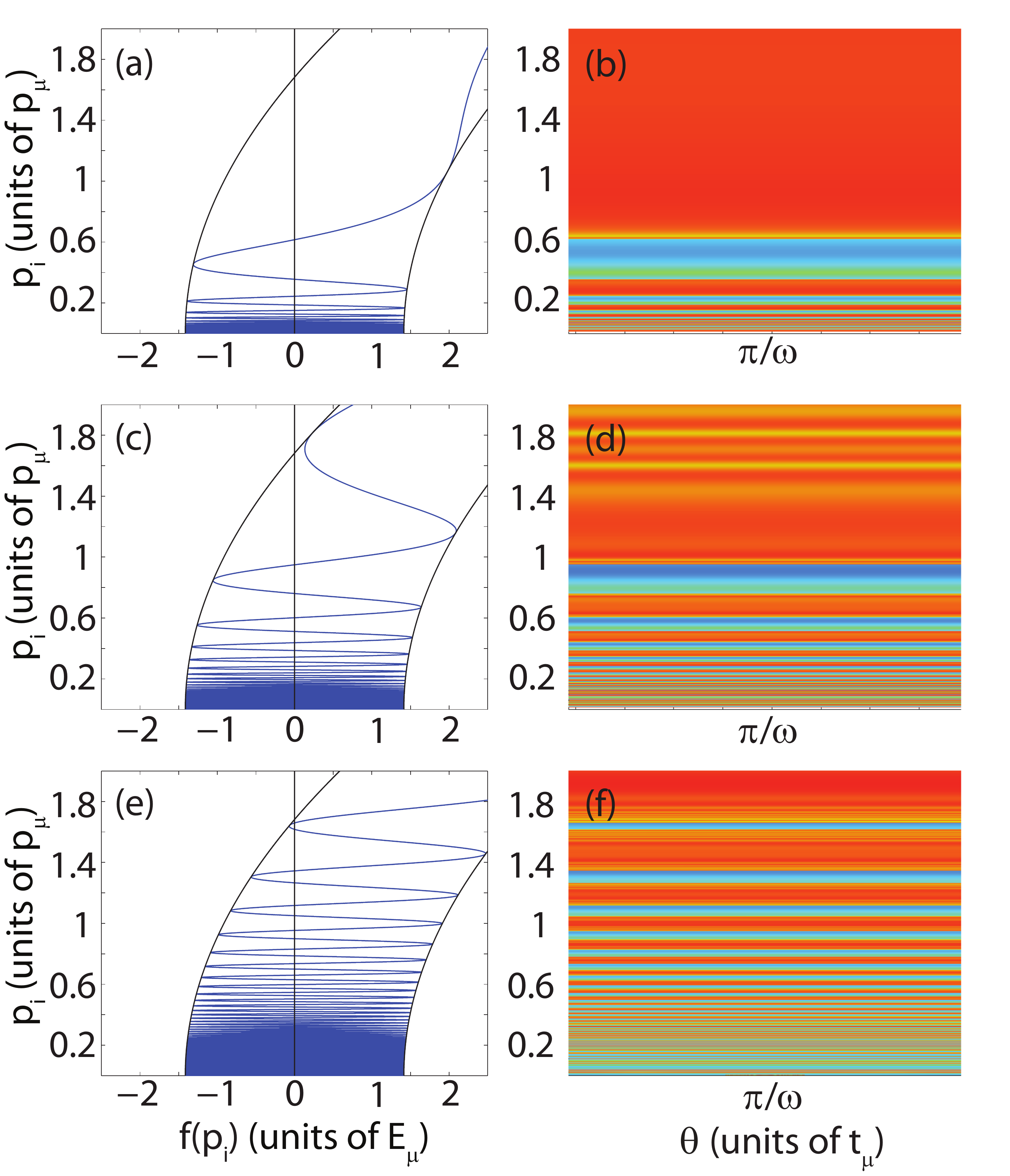}
\caption{(Color online) Increasing barrier width decreases the widths of transmission and reflection ribbons. The zeroes of the functions in the left column mark the boundaries of transmission and reflection at the chosen $\theta$. The right column shows zooms of $p_f(|p_i|,\theta)$ for the curves to the left. The top row has $\hat{x}=\sigma=1.25$, the middle row has $\hat{x}=\sigma=5$, and the bottom row has $\hat{x}=\sigma=20$ with all other barrier parameters equal.}\label{zeroes}
\end{figure}

Eq.~\eqref{fzero} is plotted for the selected barrier parameters in the left column of Fig.~\ref{zeroes} while varying $\sigma$, the barrier width. The oscillatory curve is $f\left(p_i\right)$, and the two quadratic curves are $p_i^2/2 \pm \sqrt{2}$, which bound $f\left(p_i\right)$. The threshold $p_i=2^{3/4}$ is the intersection between the left-most quadratic curve and the vertical line. The right column contains a zoom of $p_f(|p_i|,\theta)$ about $\theta=\pi/\omega$ for particles incident from the left for the respective $\sigma$. The top row represents $\hat{x}=\sigma=1.25$ (see Fig.~\ref{rbnospace}), the middle row is $\hat{x}=\sigma=5$ (see Fig.~\ref{wider}), and the bottom row has $\hat{x}=\sigma=20$. The infinite number of bands, and the reduction in their widths as the barriers get wider, are evident in these pictures.

\subsection{Separated barriers}\label{recseparate}

Inserting space between the barriers leaves many of the features of the preceding section intact, but introduces a critical difference in particle trajectories.  Previously, a particle could reflect from a barrier no more than  once. However, particles may now become trapped between the barriers for a long time, reflecting back and forth between them before finally arriving at the edge of a barrier with enough energy to transmit over it. These particle trajectories are thus very sensitive to initial conditions and the system is a model of chaotic scattering.

Numerical calculation of final momentum is performed in similar fashion as before. If a particle beginning on the left of the pump with positive momentum has enough energy to hop onto the left-hand barrier, we calculate its momentum $p_{b_1}$ and the time $t_0$ at which it reaches the end of this barrier using Eqs.~\eqref{pa} and~\eqref{ta}, respectively.  However, instead of either reflecting from the right barrier or transmitting over it, the particle instead falls off the first barrier into the region between barriers with momentum

\begin{eqnarray}
p_N=(-1)^{(N-1)}\sqrt{2\left(U_L(t_0)+\frac{p_{b_1}^2}{2}\right)},\label{prightmid}
\end{eqnarray}
with $N=1$.  The particle reaches the second barrier at time

\begin{figure}[t]
\includegraphics*[width=\columnwidth]{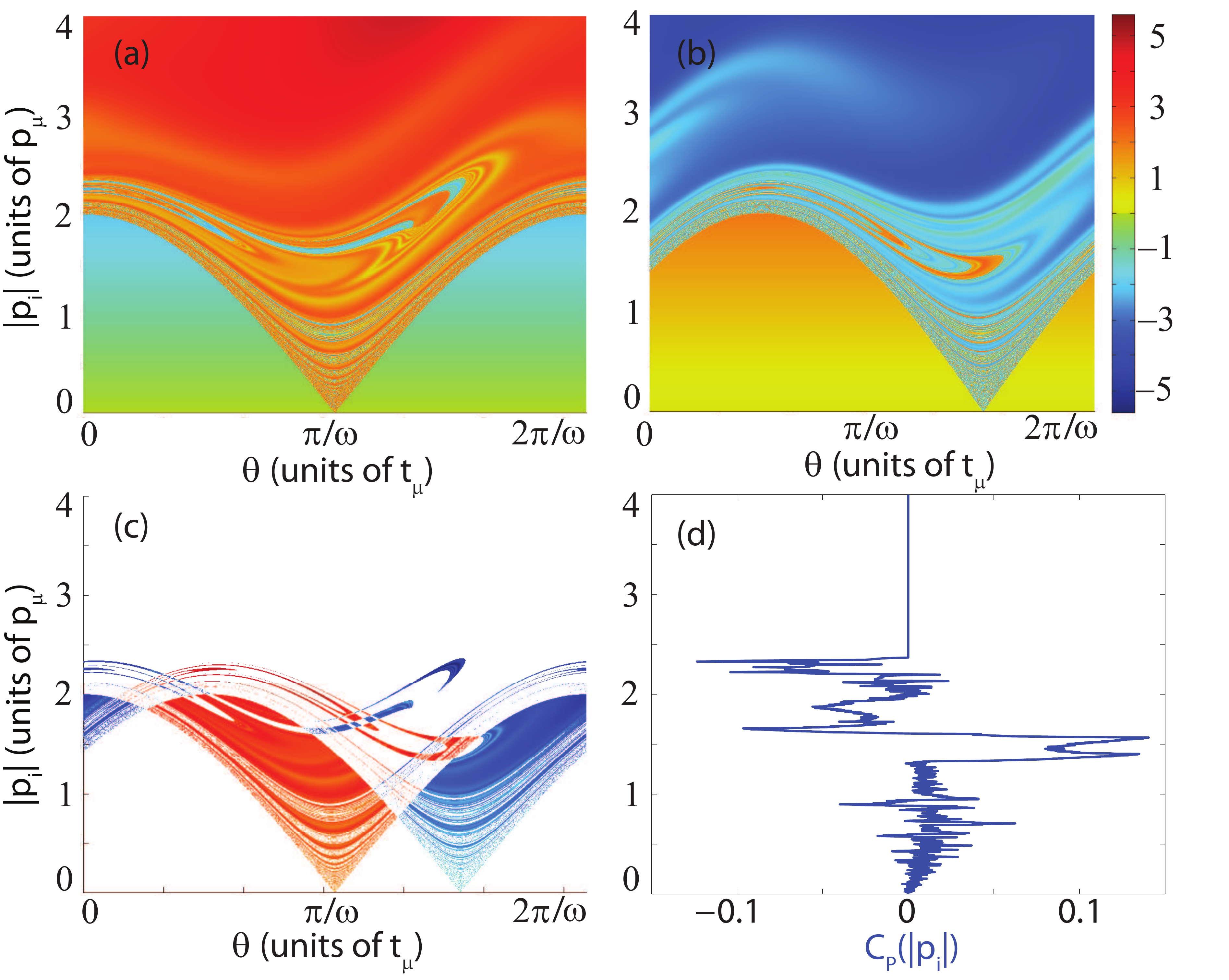}
\caption{(Color Online) . Same as Fig.~\ref{wider}, but with space between the barriers. The barriers are now centered at $\pm \hat{x} = \pm 15$, and have a distance of $d=20$ between their inner edges.}\label{rbwspace}
\end{figure}

\begin{align}
t_N&=N\frac{d}{p_N}+t_0,
\end{align}
where
\begin{align}
d&=2\hat{x}-2\sigma\label{dsep}
\end{align}
is the distance between the inner edges of the barriers.

If the height of the second barrier is greater than the particle's energy, i.e., $U_R(t_N)>p_N^2/2$, the particle reflects from the second barrier, and we increment $N$ by $1$. The index $N$ thus counts the number of trips between the barriers for each trajectory.  Each time a particle arrives at the edge of a barrier, we compare its kinetic energy $p_N^2$/2 with the height of that barrier (for odd $N$ we compare to $U_R(t_N)$, and for even $N$ we compare to $U_L(t_N)$) until it has enough energy to hop onto a barrier. Once on top of a barrier, the particle's momentum is given by

\begin{eqnarray}
p_{b_2}=\pm\sqrt{2\left(\frac{p_N^2}{2}-U_{R,L}(t_N)\right)},
\end{eqnarray}
where $p_{b_2}$ is positive for odd $N$, and negative for even $N$. The particle then falls off the second barrier at time

\begin{eqnarray}
t_b=\frac{2\sigma}{p_{b2}}+t_N,
\end{eqnarray}
with final momentum

\begin{eqnarray}
p_f=\pm\sqrt{2\left(\frac{1}{2}p_{b_2}^2+U_{R,L}(t_b)\right)},
\end{eqnarray}
where $p_f>0$ if $p_{b_2}>0$, and $p_f<0$ if $p_{b_2}<0$. The calculation is similar for particles approaching the barriers from the right.

\begin{figure}[t]
\includegraphics*[width=\columnwidth]{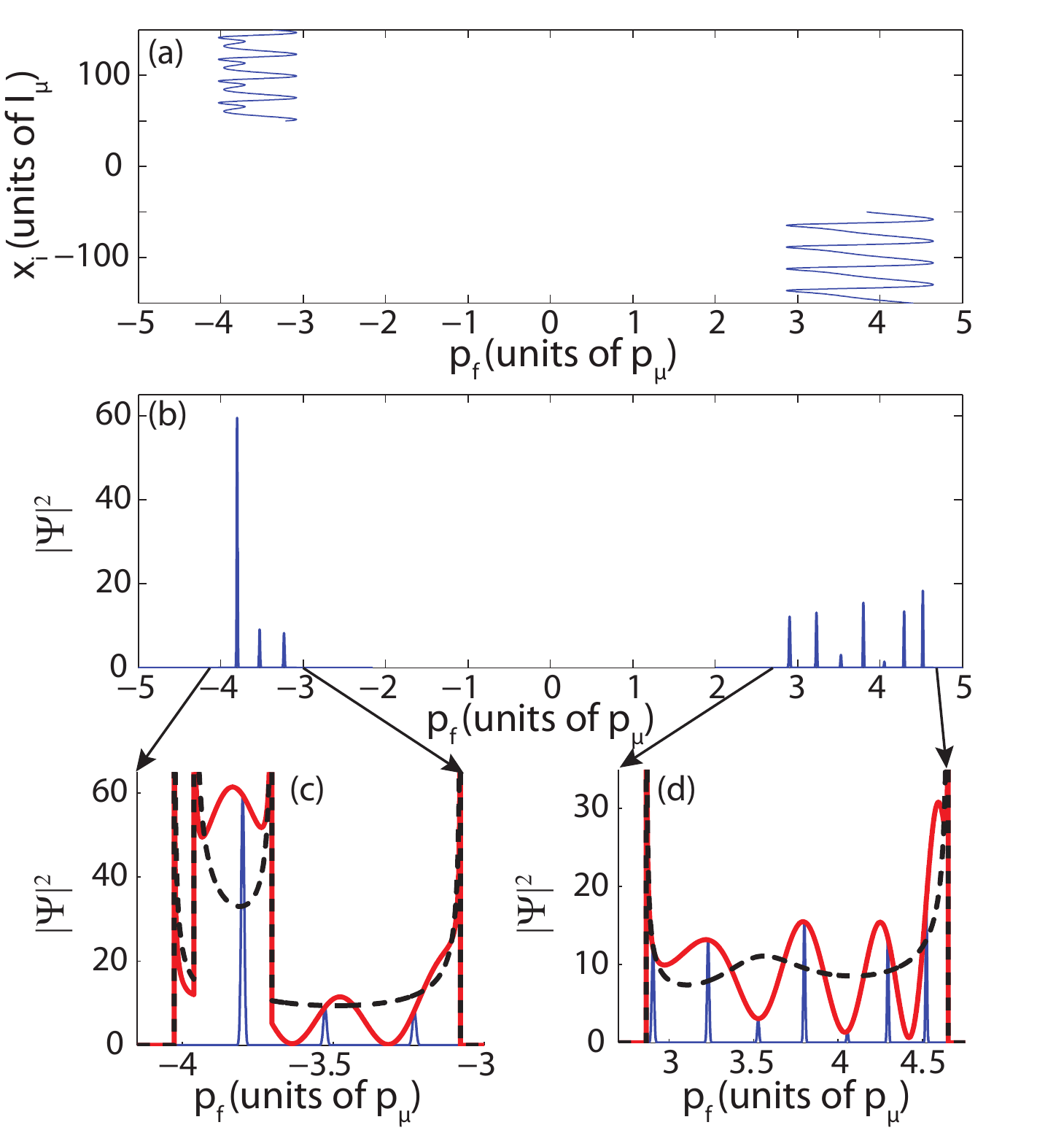}
\caption{(Color Online) Classical and primitive semiclassical final momentum probabilities for the separated barriers in Fig.~\ref{rbwspace}. All curves are as described in Fig.~\ref{rectrect}}\label{scspace}
\end{figure}

Figure~\ref{rbwspace} shows results for a pump with potentials given by Eqs.~\eqref{recleft} and~\eqref{recright} with $\hat{x}=15$, $\sigma=5$, $\hat{U}=1$, $\alpha=1$, $\omega=1$, and $\phi=3\pi/2$. This pump is the same as the one from Fig.~\ref{wider}, except the inner edges of the barriers are now separated by a distance $d=20$. Similar to the effect of making the barriers wider, inserting space between the barriers affects the ribbons of transmission and reflection for particles approaching the barriers from both sides. The width ($\Delta |p_i|$) of the ribbons decays more quickly as $|p_i| \rightarrow 0$ for a given $\theta$. Consequently, the width $\Delta |p_i|$ for regions of large net particle transport  is smaller. The magnitude of $C_P(|p_i|)$ has decreased in this example (although increasing the space between the barriers can also cause it to increase). Predicting the effect of increasing barrier separation on the magnitude of fractional particle transport is not possible without detailed calculations.

Fig.~\ref{scspace} shows $x_i(p_f)$ for two particle packets which approach the barriers from opposite sides with the same initial energy ($p_i=\pm $3.8). The initial packets are described by Eq.~\eqref{rectenv} with $-x_c=\mp 450$ and $\beta=100$. Their initial energy is large enough such that all particles transmit over both barriers. Particles approaching from the left are scattered to a larger range of $p_f$ than those approaching from the right. This results in more peaks for $p_f>0$ in the semiclassical probability density $P^{SC}_f(p_f)$ shown in Fig.~\ref{scspace}(b). Figs.~\ref{scspace}(c) and (d) show expansions of Fig.~\ref{scspace}(b) (note the different scales). This calculation has not been extended into the classically-forbidden regions. In Fig.~\ref{scspace}(c), the classical probability density $P^C(p_f)$ (dashed curve) and single-cycle semiclassical probability density (oscillatory curve) $P^{SC}_s(p_f)$ diverge at each of the four turning points of $x_i(p_f)$. Interference among the four branches of $x_i(p_f)$ within one cycle causes $P^{SC}_s(p_f)$ to be larger than $P^{C}(p_f)$ at the location of the largest peak, $p_f \approx -3.53$. $P^{SC}_s(p_f)$ and $P^{C}(p_f)$ are scaled (multiplied by the same constant) in order to be plotted with $P^{SC}_f(p_f)$ (sharply-peaked curve), and the relative heights in the Floquet peaks can be seen to align closely with the discrete values of $P^{SC}_s(p_f)$ at momenta corresponding the Floquet energies.

Analysis of transmission and reflection ribbons is more difficult in the present case because of the possibility of multiple reflections between the barriers. However, we can gain insight by analyzing criteria for particles which directly transmit past both barriers with no reflection. For particles approaching from the left with $p_i>0$ and arriving at the left-hand barrier at $\theta=\pi/\omega$ (i.e. when the height of the left barrier is zero), the condition for direct transmission past both barriers is

\begin{eqnarray}
f_d\left(p_i,t_0\right)=p_i^2/2+U_L(t_0)-U_R(t_N)>0\label{eqnhere}
\end{eqnarray}
with $N=1$, where the subscript $d$ is the distance between the inner edges of the barriers (see Eq.~\eqref{dsep}). Eq.~\eqref{eqnhere} reduces to
\begin{align}
f_d\left(p_i\right)=\frac{p_i^2}{2}-2\sin\left(\frac{\hat{x}-\sigma}{p_N}+\frac{3\pi}{4}\right)
 \cos\left(\frac{\hat{x}-\sigma}{p_N}+\frac{\pi}{4}+\frac{2\sigma}{p_i}\right)\label{f2}
\end{align}
for our selected barrier parameters.
\begin{figure}[t]
\includegraphics*[width=\columnwidth]{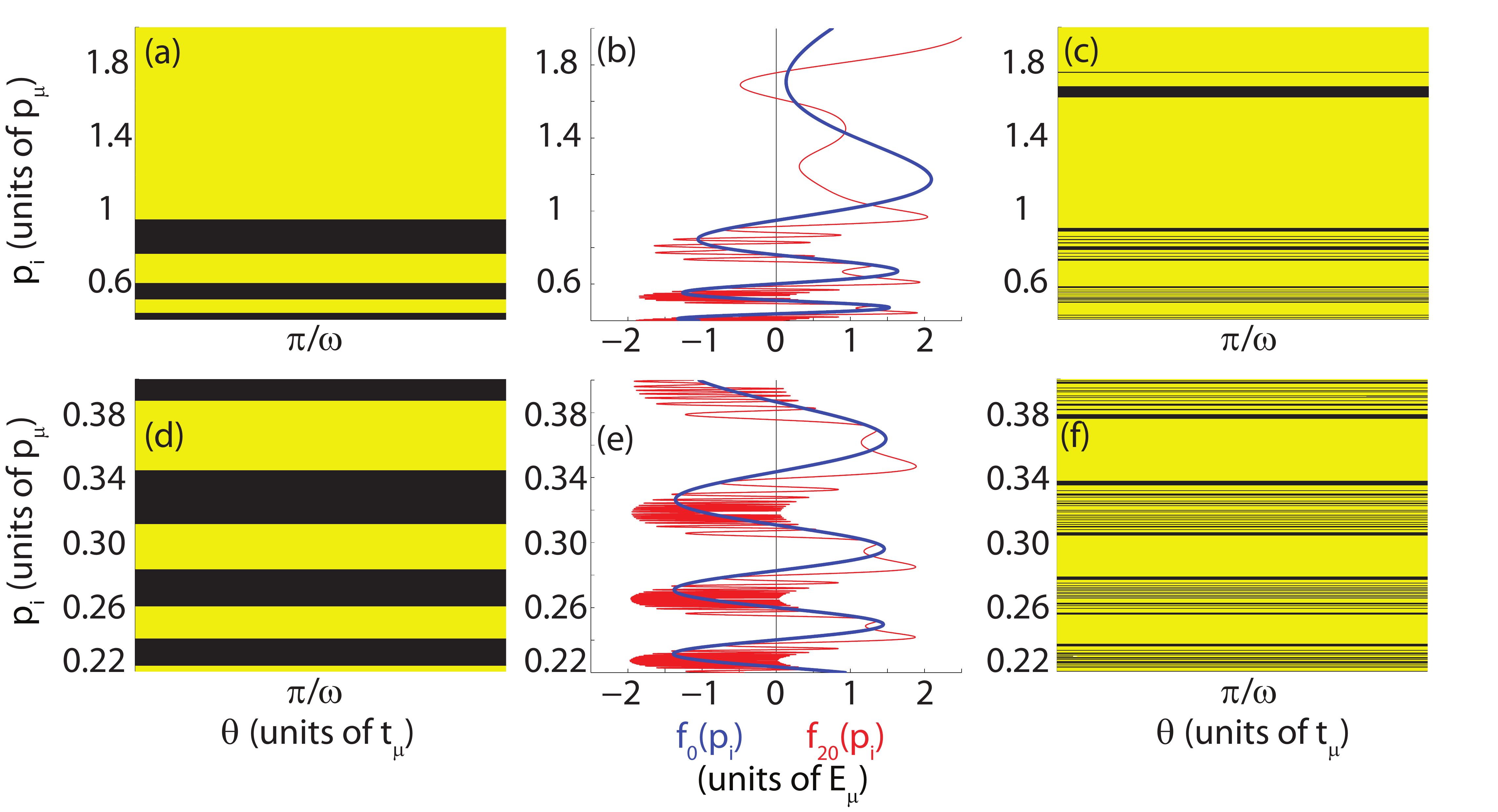}
\caption{(Color Online) Left column: Zooms of Fig.~\ref{wider}(a) about $\theta=\pi/\omega$ with color schemed changed. Black represents reflection; yellow represents transmission. Right column: Zooms of Fig.~\ref{rbwspace}(a) about $\theta=\pi/\omega$. The zeroes of the thick (blue) oscillatory curve in the middle column mark the boundaries of transmission and reflection seen in the left column. When the thin (red) oscillatory curve is positive, particles in the right column transmit, but this function does not reveal all transmission ribbons in the right column.}\label{f2zero}
\end{figure}

The left and right columns of Fig.~\ref{f2zero} show zooms of Figs.~\ref{wider}(a) ($\hat{x}=\sigma=5$) and \ref{rbwspace}(a) ($\hat{x}=15$ and $\sigma=5$), respectively, about $\theta=\pi/\omega$. The color scheme has been changed to enhance visibility; black ribbons represent reflection and the lighter (yellow online) ribbons represent transmission. The middle column of Fig.~\ref{f2zero} shows $f_d\left(p_i\right)$ for these two pumps. The thick (blue online) curve is $f_0\left(p_i\right)$ ($d=0$), and is the same curve seen in Fig.~\ref{zeroes}. Its zeroes mark the boundaries of transmission and reflection in the left column. The thin curve (red online) is $f_{20}\left(p_i\right)$ ($d=20$), and corresponds to the right column.

When $f_{20}\left(p_i\right)>0$, particles arriving at the pump in the right column transmit, and the ribbon in the right column is the light color (yellow online). When $f_{20}\left(p_i\right)<0$, the particle reflects from the right-hand barrier, and its ultimate fate is unspecified. It is evident that $f_{20}\left(p_i\right)$ oscillates more rapidly than $f_0\left(p_i\right)$. Consequently, regions which reflect when the barriers touch are split into multiple transmission and reflection ribbons as the barriers are moved apart. This is illustrated in the present case in regions where $f_{0}\left(p_i\right)<0$ and $f_{20}\left(p_i\right)>0$. For each region in which $f_{0}\left(p_i\right)$ is negative, there is a reflection ribbon in the left column. However, in each such region, $f_{20}\left(p_i\right)$ oscillates through zero many times, and each positive segment of $f_{20}\left(p_i\right)$ represents a transmission ribbon in the right column. Increasing the barrier separation distance thus creates many transmission and reflection ribbons in regions where there is only pure reflection when the barriers touch. The minimum $p_i$ above which all particles arriving at $\theta=\pi/\omega$ transmit has also been greatly increased by moving the barriers apart. With no barrier separation, this $p_i \approx 0.95$, but increasing $\hat{x}$ to $15$ increases this minimum to $p_i \approx 1.76$. In each case, there are an infinite number of ribbons as $p_i\rightarrow0$.

This level of analysis predicts only the outcome of each particle's first arrival at the right-hand barrier. What happens after that is ``left as an exercise for the reader.''

The effects of increasing barrier width and separation can be summarized as follows. Increasing the width causes more transmission and reflection ribbons below arbitrary $|p_i|$, up to a maximum $|p_i|$ above which all particles will transmit for a given $\theta$. The width of the ribbons (in terms of $\Delta |p_i|$) decays more slowly as $|p_i|\rightarrow 0$ for wider barriers. Ribbons produced by particles incident upon barriers with no separation are split into multiple ribbons by moving the barriers apart. Increasing barrier separation can also allow particles of much higher $|p_i|$ to reflect for a given $\theta$. Increasing either of these parameters causes the widths ($\Delta |p_i|$) of regions in which there is significant fractional particle transport to decrease, although its magnitude is not systematically changed. Predictions on fractional transport are highly sensitive to the choice of initial conditions and parameters, and do not display any obvious pattern. Therefore, general predictions beyond what we have mentioned cannot be made without detailed calculations specific to a configuration and choice of parameters.

\section{\label{gaussbarr}Gaussian Barriers}
While rectangular barriers provide a simplified model that addresses the essential pumping physics, Gaussian barriers are more likely to be used in experimental implementations using laser-based optical dipole barriers for ultracold atoms.  In this section, we examine a turnstile pump (such as those in Section~\ref{symm}) with Gaussian potentials described by Eqs.~\eqref{gaussleft} and~\eqref{gaussright} with $\hat{U}_L=\hat{U}_R=1$, $\alpha=1$, $\omega=1$, $\phi=3\pi/2$, $\sigma=2.5/(2\sqrt{2\ln2})$, and $\hat{x}=3.75$. Both barriers oscillate at the same frequency, but not in phase with one another. As in the previous section, particle trajectories for this type of pump are classically chaotic.

\begin{figure}[t]
\includegraphics*[width=\columnwidth]{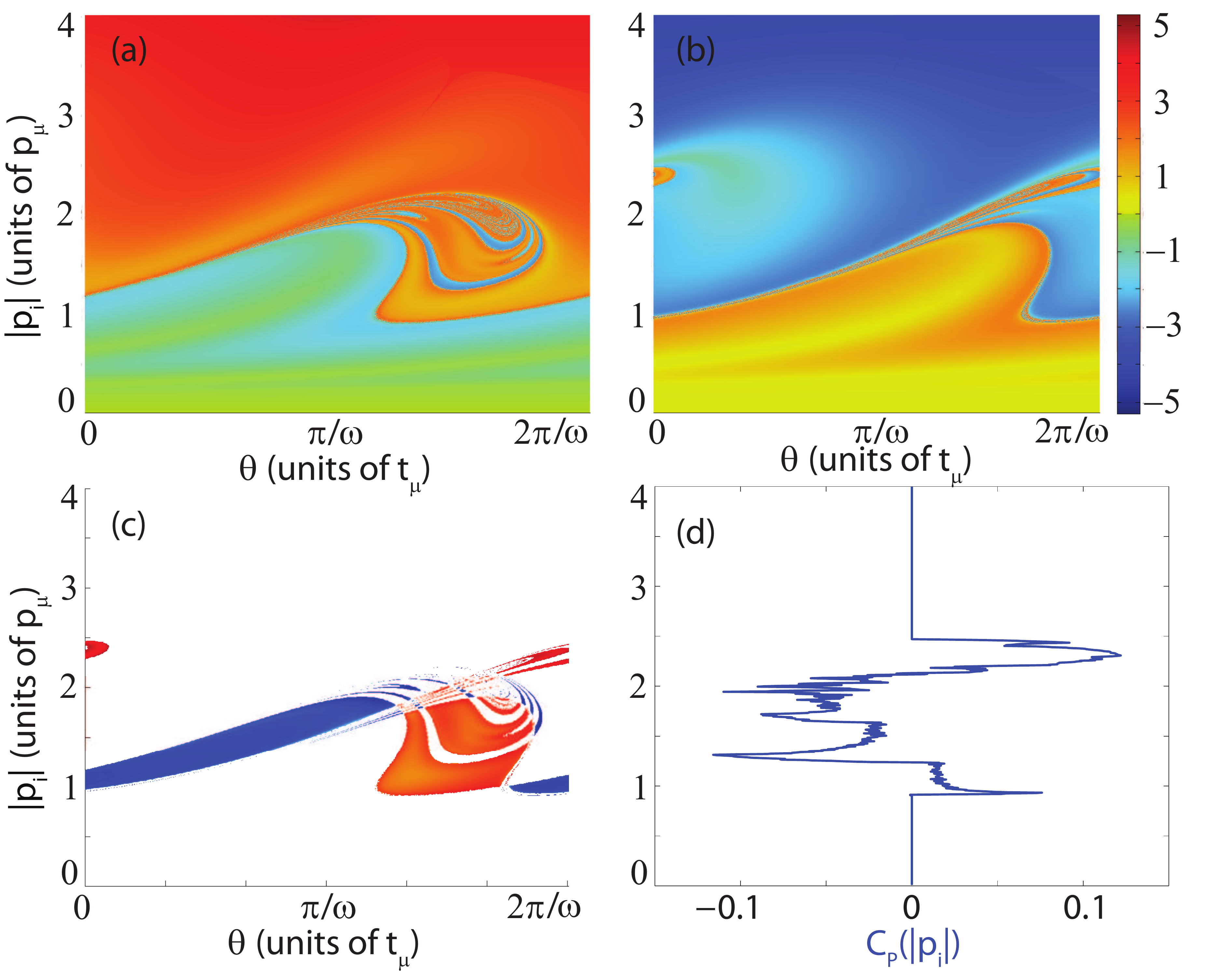}
\caption{(Color Online) Same as Fig.~\ref{rbwspace}, but for Gaussian barriers.}\label{redo}
\end{figure}

Fig.~\ref{redo}(a) and (b) show $p_f(|p_i|,\theta)$ for particles incident upon the barriers from the left and right, respectively. Unlike the previous cases with rectangular barriers, there is a minimum $|p_i|$ below which there is no particle transmission. As particles approach Gaussian-shaped repulsive barriers, they lose momentum, resulting in a minimum initial energy required to transmit past the first barrier encountered. In this case, all particles with $|p_i|\lesssim 0.90$ reflect from the first barrier. Different types of structure can be seen in $p_f(|p_i|,\theta)$ than for rectangular barriers, but qualitative features remain. The regions in which striping can be seen indicate particle trajectories which are temporarily trapped between the barriers before finally transmitting or reflecting. The lobe with significant striping seen in Fig.~\ref{redo}(a) is much wider than the narrow one seen in Fig.~\ref{redo}(b) in the range $1.75 \lesssim |p_i| \lesssim 2.25$, indicating that particles approaching from the left in this energy range are much more likely to become temporarily trapped between the barriers than those approaching from the right with equal energy. However, above this range, all particles approaching from the left transmit, while some approaching from the right are trapped between the barriers until $|p_i|\gtrsim 2.5$. The complete description of particle transport through the barrier region lies outside the scope of this paper, but a detailed topological analysis is given in \cite{ByrdDelos}.

 Fig.~\ref{redo}(c) shows $(|p_i|,\theta)$ for which both particles scatter to the left (blue online) or right (red online). Whereas the previous cases with rectangular barriers have structure as $|p_i|\rightarrow 0$, no structure is present in this region for Gaussian barriers because of the nonzero minimum $|p_i|$ required to transmit past the first barrier. Fig.~\ref{redo}(d) shows $C_P(|p_i|)$. The vertical region in this curve below $|p_i|\lesssim 0.90$ corresponds to the region in which all particles directly reflect. Above this range, fractional particle transport occurs in both directions until $|p_i|$ is large enough for all particles to transmit past both barriers.

\begin{figure}[t]
\includegraphics*[width=\columnwidth]{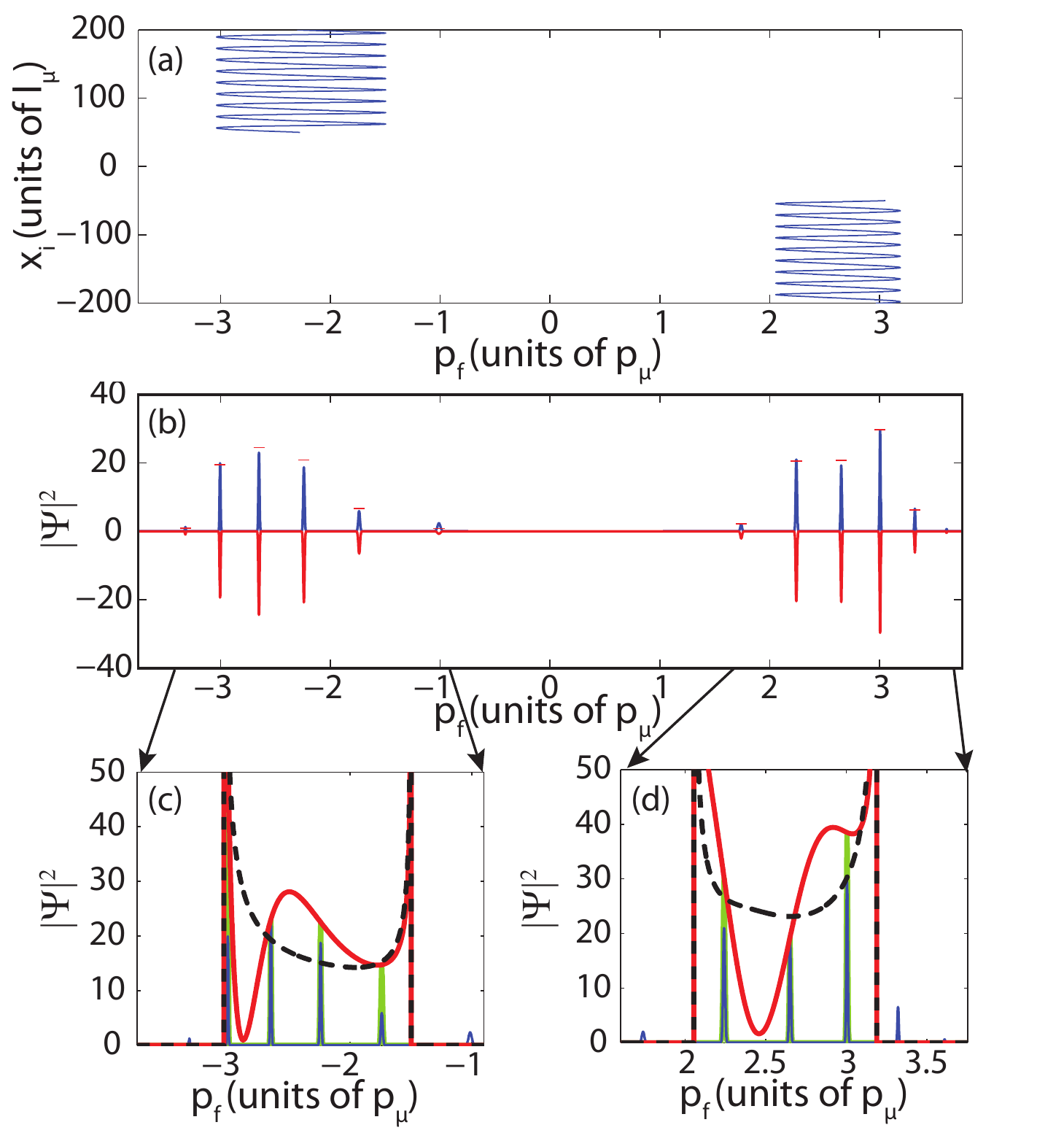}
\caption{(Color Online) (a) $p_f(x_i)$ for particles approaching a pump with two oscillating Gaussian barriers from both sides. (b) Final uniform semiclassical momentum probability $P^{SC}_f(p_f)$ (plotted upward, blue), and quantum mechanical final momentum probability $P^Q_f(p_f)$ (plotted downward, red), for the particles in (a). (c) and (d) $P^C(p_f)$ (dashed curve), $P^{SC}_s(p_f)$ (oscillatory curve, red online), $P_P^{SC}(p_f)$ (green online) and $P^{SC}_f(p_f)$ (blue online) for the particles in (a). See the text for a discussion of these functions.}\label{gauss2osc}
\end{figure}

Figure~\ref{gauss2osc} shows classical, semiclassical, and quantum-mechanical comparisons for two packets of particles approaching the barriers from opposite directions with  $p_i=\pm  2.65$. The initial packets are decribed by Eq.~\eqref{rectenv} with $-x_c=\mp 450$ and $\beta=100$.

Figure~\ref{gauss2osc}(a) shows classical initial position as a function of final momentum for particles approaching the barriers from both sides. All particles have enough energy to transmit over both barriers. Particles incident upon the barriers from the right scatter to a larger range of $\Delta p_f$ than those approaching from the left.

Figure~\ref{gauss2osc}(b) shows $P^{SC}_f(p_f)$, the uniform semiclassical final momentum probability density (plotted upward), and $P^Q_f(p_f)$, the quantum mechanical final momentum probability (plotted downward). The uniform semiclassical calculation has been repaired near turning points of $p_f(x_i)$, where the primitive form is divergent, and has been extended into classically-forbidden regions. $P^Q_f(p_f)$ has been mirrored about the amplitude axis for ease in comparing the two calculations. The horizontal lines (red online) in the upper half-plane are the heights of the peaks in $P^Q_f(p_f)$, and are plotted to allow one to compare the calculations more easily. Very good agreement between the two methods is evident.

Figures~\ref{gauss2osc}(c) and (d) show the classical probability density $P^c(p_f)$ (dashed curve), and the primitive semiclassical single-cycle probability density $P^{SC}_s(p_f)$ (thick oscillatory  curve, red online). The two sharply-peaked functions are the primitive semiclassical probability density $P^{SC}_p(p_f)$, given by summing Eq.~\eqref{scpsi} for all branches (lighter peaked curve, green online), and the uniform semiclassical probability density $P^{SC}_f(p_f)$ (darker peaked curve, blue online). The functions $P^C(p_f)$ and $P^{SC}_s(p_f)$ are scaled (multiplied by the same constant). The function $P^{SC}_p(p_f)$ (green online) takes on discrete values of the curve $P^{SC}_s(p_f)$ (red online) at momenta satisfying $E_n=E_i+n\hbar\omega$, showing that the single-cycle probability governs the relative heights of the Floquet peaks (the single-cycle probability shown is the primitive form).

The classical dynamics underlying the quantum treatment are therefore necessary to fully understand the quantum mechanical result. While quantum theory tells us that the density will be peaked at momenta satisfying $E_n=E_i+n\hbar\omega$, it does not tell us the range of $n$ for which the peaks will be of appreciable height. The final momentum region in which particles are classically scattered governs the range of $n$ for which the quantum result yields large peaks. Quantum theory also does not indicate why some momentum states are more highly populated than others, but semiclassical tools give an intuitive explanation for that.

The double barrier turnstile pump might be viewed (with some caution) as a momentum-space interferometer. In this picture, each oscillating barrier acts as a multichannel beamsplitter which takes an incoming planewave and transforms it into a superposition of outgoing planewaves with different momenta (with energies $E_n=E_i+n\hbar\omega$). In a pure transmission case (such as Fig.~\ref{gauss2osc}), the first barrier produces multiple planewave states, and then the second barrier mixes these and produces additional planewave states. In this way a turnstile pump may be viewed as a discrete multipath momentum space interferometer. However, this description cannot be accurate if the barriers are not well-separated. The barriers must be sufficiently far apart that the configuration-space wave function in the region between them is approximately a superposition of plane waves, but not so far apart that packets associated with different Floquet states have separated.

\section{\label{conc}Conclusion}

In summary, we have defined and described ballistic atom pumps, showing that for finite ranges of initial particle energies, such systems can create net particle transport in either direction.  The direction of particle pumping  is highly sensitive to barrier parameters and to the initial energy of the particles.  It is not possible to predict the direction or magnitude of particle pumping without detailed calculations.

If tunneling can be neglected, diode pumps--which only allow net transport in one direction for particles below a certain initial energy--can be constructed. At sufficiently high incident particle energies, these diodes only allow net particle transport in the opposite direction.

We have studied these pumps classically, semiclassically, and quantum mechanically. While classical theory gives a slowly-varying final momentum probability for scattered particles, quantum theory yields final momentum probabilities sharply-peaked at momenta satisfying $E_n=E_i+n\hbar\omega$. The range of $n$ for which there are appreciable peaks is governed by the underlying classically-allowed momentum range of scattered particles. Semiclassical theory gives an intuitive explanation for the relative heights of the peaks, and agrees well with the quantum description.

\section{Acknowledgements}
T. A. B. and M. K. I. contributed equally to this paper. K. K. D. acknowledges support of the NSF under Grant No. PHY-1313871.  K. A. M. acknowledges NSF support from Grant No. PHY-0748828. J. B. D. and T. A. B. acknowledge NSF support via Grant No. PHY-1068344.


\end{document}